\documentclass[11pt]{amsart}
\usepackage{fullpage}
\usepackage{amsmath,amsthm,amssymb,booktabs}
\usepackage{tabularx,epsfig,graphicx,url,dcolumn,etoolbox}
\usepackage{natbib} 

\title{A comparison of partisan-gerrymandering measures}

\ifdef{\SUBMIT}{}
{
\author{Gregory S. Warrington}
\address{Department of Mathematics \& Statistics, University of Vermont,16 Colchester Ave., Burlington, VT 05401, USA}
\email{gswarrin@uvm.edu}
}

\date{}

\begin{document}

\ifdef{\SUBMIT}
      {
        \baselineskip 24pt
        }{}

\newcommand{\bp}{\boldsymbol{p}}
\newcommand{\waste}{w}
\newcommand{\barp}{\overline{p}}

\newcommand{\scale}[1]{\mathrm{#1}}
\newcommand{\noscale}[1]{\mathrm{#1}}
\newcommand{\exelec}{\mathcal{E}_0}

\newcolumntype{L}{D{.}{.}{2,2}}

\newcommand{\taugap}{{EG}_{\tau}^1}
\newcommand{\lambdaeg}[1]{{{EG}^{#1}}}
\newcommand{\eg}{{\lambdaeg{1}}}
\newcommand{\dg}{{\lambdaeg{2}}}
\newcommand{\sg}{{WE}}
\newcommand{\lossg}{{\lambdaeg{0}}}
\newcommand{\vcg}{{EG}_v}
\newcommand{\va}{\vcg^1}
\newcommand{\vb}{\vcg^2}

\newcommand{\mm}{MM}
\newcommand{\bias}{{Bias}}
\newcommand{\biasO}{{Bias'}}

\newcommand{\dec}{{Dec}}
\newcommand{\bdec}{{Dec'}}

\newcommand{\lm}{{Lop}}
\newcommand{\evw}{{EVW}}

\newcommand{\elecmac}[1]{\textsc{#1}}
\newcommand{\elecA}{\elecmac{1-Prop}}
\newcommand{\elecB}{\elecmac{2-Prop}}
\newcommand{\elecC}{\elecmac{3-Prop}}
\newcommand{\elecD}{\elecmac{Sweep}}
\newcommand{\elecE}{\elecmac{Comp}}
\newcommand{\elecF}{\elecmac{Comp'}}
\newcommand{\elecG}{\elecmac{Uncomp}}
\newcommand{\elecH}{\elecmac{V-Uncomp}}
\newcommand{\elecI}{\elecmac{Cubic}}
\newcommand{\elecJ}{\elecmac{Anti-M}}
\newcommand{\elecK}{\elecmac{Classic}}
\newcommand{\elecL}{\elecmac{Inverted}}

\newcommand*\rot{\rotatebox{90}}


\newcommand{\mycapa}{Illustration of a $5$-district election $\mathcal{E}_0$
    with $\bp = (0.35,0.4,0.45,0.6,0.8)$ and $\barp = 0.52$. We note
    for reference that $k=3$ and $k'=2$ (see equation~\eqref{eq:bp}).}

\newcommand{\mycapb}{Definition of the declination measure.}

\newcommand{\mycapc}{Kernel density plots for each rescaled
  measure. Plots are split into three groups for clarity. We have
  overlaid a standard Gaussian distribution on each plot for
  reference. The equal vote weights measure is omitted since it is
  numerically equivalent to the mean-median difference.}

\newcommand{\mycapd}{Scatter plots corresponding to various pairs of
    measures. Each plot consists of the $1,166$ elections with at least
    seven districts and for which each party wins at least one seat.}

\newcommand{\mycape}{Most extreme outliers among balanced elections for
    $\scale{\dg}$, $\scale{\eg}$, $\scale{\lossg}$, $\scale{\va}$ and
    $\scale{\vb}$. In this and later figures, values of measures
    listed are for the rescaled versions.}

\newcommand{\mycapf}{Most extreme outliers among all elections for
    $\scale{\dg}$, $\scale{\eg}$, $\scale{\lossg}$, $\scale{\va}$ and $\scale{\vb}$.}

\newcommand{\mycapff}{Most extreme outliers among balanced elections (top row)
    and all elections (bottom row) for $\scale{\taugap}$.}

\newcommand{\mycapg}{Most extreme outliers among balanced elections (top row)
    and all elections (bottom row) for the declination,
    $\scale{\dec}$.}

\newcommand{\mycaph}{Most extreme outliers among balanced elections (top row)
    and all elections (bottom row) for the buffered declination,
    $\scale{\bdec}$.}

\newcommand{\mycapi}{Most extreme outliers among balanced elections (top row)
    and all elections (bottom row) for the mean-median difference,
    $\scale{\mm}$.}

\newcommand{\mycapj}{Most extreme outliers among all elections for
    $\scale{\evw}$. The same three elections are identified among the
    subpopulation of balanced elections.}

\newcommand{\mycapk}{Most extreme outliers among balanced elections (top row)
    and all elections (bottom row) for partisan bias, $\scale{\bias}$.}

\newcommand{\mycapkii}{Most extreme outliers among balanced elections (top row)
    and all elections (bottom row) for partisan bias, $\scale{\biasO}$.}

\newcommand{\mycapl}{Most extreme outliers among balanced elections (top row)
    and all elections (bottom row) for the lopsided-means test,
    $\scale{\lm}$. Elections shown are restricted to those for which
    the Student $t$-test implies significance.}

\newcommand{\mycapm}{Twelve hypothetical elections shown along with mean
    Democratic vote fraction. These elections are evaluated by the
    various scaled measures in Table~\ref{tab:hypo}.}

\newcommand{\mycapn}{Greatest pairwise disagreements between $\scale{\dec}$ and
    $\scale{\bias}$ among balanced elections (top row) and all
    elections (bottom row).}

\newcommand{\mycapo}{Greatest pairwise disagreements between $\scale{\dec}$ and
    $\scale{\mm}$ among balanced elections (top row) and all
    elections (bottom row).}

\newcommand{\mycapp}{Greatest pairwise disagreements between $\scale{\eg}$ and
    $\scale{\dec}$ among balanced elections (top row) and all
    elections (bottom row).}

\newcommand{\mycapq}{Greatest pairwise disagreements between $\scale{\eg}$ and
    $\scale{\bias}$ among balanced elections (top row) and all
    elections (bottom row).}

\newcommand{\mycapr}{Greatest pairwise disagreements between $\scale{\eg}$ and
    $\scale{\mm}$ among balanced elections (top row) and all
    elections (bottom row).}

\newcommand{\mycaps}{Greatest pairwise disagreements between $\scale{\mm}$ and
    $\scale{\bias}$ among balanced elections (top row) and all
    elections (bottom row).}

\newcommand{\mycapt}{Elections on which measures come to the least consensus.}

\newcommand{\mycapu}{Representative elections for various levels of measure
    value ranging from $0$ to $3$ standard deviations away from the
    respective mean. Elections were drawn from the population of
    balanced elections.}

\newcommand{\mycapv}{Example of two elections with the same statewide Democratic
    average. The buffered declination for the North Carolina election
    indicates a mildly pro-Republican advantage with the Republicans
    winning three seats while the (shifted) Virginia election
    indicates a mildly pro-Democratic advantage with the Republicans winning
    four seats.}

\newcommand{\mycapsens}{Exploration of the sensitivity of measures to
  variation in statewide vote share from -5\% to +5\% for four
  hypothetical elections.}


\begin{abstract}
  We compare and contrast fourteen measures that have been proposed
  for the purpose of quantifying partisan gerrymandering. We consider
  measures that, rather than examining the shapes of districts,
  utilize only the partisan vote distribution among districts. The
  measures considered are two versions of partisan bias; the
  efficiency gap and several of its variants; the mean-median
  difference and the equal vote weight standard; the declination and
  one variant; and the lopsided-means test. Our primary means of
  evaluating these measures is a suite of hypothetical elections we
  classify from the start as fair or unfair. We conclude that the
  declination is the most successful measure in terms of avoiding
  false positives and false negatives on the elections considered. We
  include in an appendix the most extreme outliers for each measure
  among historical congressional and state legislative elections.
\end{abstract}

\maketitle

\section{Introduction}

A partisan gerrymander occurs when one political party illicitly draws
electoral district lines to its own advantage. A number of measures
that are based on how voters of each party are distributed among
districts have been proposed as aids to identifying partisan
gerrymanders. Unfortunately, there is no consensus as to which
of these measures works the best.

There have been several efforts to compare some of the proposed
measures. In ~\citep{MSII}, the authors explore the extent to which
each of a number of measures adheres to various desiderata. These
analyses are performed by considering historical and simulated
elections in the aggregate\footnote{In this article, ``election''
  effectively refers to the results of a given election as encoded in
  the collection of Democratic vote shares across districts. For
  historical elections, these values are, of course, affected by
  everything from the overall electoral climate to the district plan
  in use to the rules governing the
  election.}. In~\citep{McDonald-two}, the authors consider an
overlapping set of measures, but do so with different evaluative
criteria applied in the context of two specific
elections. In~\citep{Nagle1,Nagle2}, a number of measures are compared
and contrasted, both on theoretical grounds and in the context of a
handful of specific elections. We refer the reader to~\citep{Tapp} for
an expository review of many of the ``wasted vote''-derived measures.

In this article we take a different tack to that taken in the
aforementioned studies. Rather than focusing on just one or two
elections or on elections in the aggregate, we focus on a dozen
hypothetical elections, each selected so as to highlight various
aspects of both fair and of unfair elections. Additionally, rather
than evaluate elections based on abstract utility functions, we ground
our evaluations in specific classes of elections that we believe should be
considered either fair or unfair. Of course, our classification of
each of these hypothetical elections is ultimately open to debate:
there is no gold standard to which we can appeal. Notwithstanding this
unavoidable truth, we aim to present our results in a way that is
illuminating even if the reader rejects both of the postulates upon
which we base our classifications.

The measures we consider in this article are closely related in that
they all look at the partisan vote by district. Most are closely tied
to the \emph{seats-votes curve}, that elusive function by which votes
are turned into seats. But the measures vary widely in how much
scrutiny they have been subjected to. The efficiency gap, explored
extensively in~\citep{M-S}, has achieved prominence through its usage
in \emph{Whitford \emph{v.}\ Gill}~\citep{Wisconsin} and has become the
de facto standard for partisan gerrymandering measures. In spite of
this success, or perhaps because of it, it has been the subject of
numerous critiques (see, e.g.,
~\citep{Griesbach,cho-upenn,Cover,MoonMira,Veomett}). On the other hand
we have measures such as the winning efficiency, mentioned almost in
passing in~\citep{cho-upenn}, for which there has been essentially no
study.

We come to several conclusions from our investigations of these
measures. First, none of the ``wasted-votes'' variations of the
efficiency gap appear to markedly improve upon it. Second, two of the
other most widely cited measures, partisan bias and the mean-median
difference, struggle in a number of ways. Third, the recently proposed
declination measure performs robustly on the entire slate of elections
we consider in this article and is therefore our recommendation for a
general-purpose measure.

\subsection{Issues not addressed in this article}
\label{sec:issues}

Measures based on the vote distribution are far from being the only
technique for quantifying or investigating partisan
gerrymanders. Irregularly shaped districts have long been viewed as
the \emph{sine qua non} of gerrymanders. A number of compactness
metrics have been developed to quantify these irregularities. Even
though compactness metrics are undeniably important, we view them
as a complementary class of measures and do not consider them in this
article.

Computer simulations have proved to be increasingly useful in
investigating gerrymandering. Much of this utility stems from the
ability to generate a baseline for what properties a fair district
plan should have. A baseline generated in this manner can take into
account specifics such as local geography, county boundaries and how
partisans are distributed geographically. Simulations can also be used
to investigate measures themselves (see, e.g.,~\citep{chatterjee}). As
with the compactness measures, we consider computer simulations to be
outside the scope of this article, in spite of their status as an
important technique for quantifying gerrymandering.

We assume in this study that, for a given election, turnout is equal
for each district. A more general analysis would consider the very
real complication that turnout can vary widely. We have chosen for
several reasons to not consider this more general setup. Primarily, we
do so because equal turnout is the simplest case. If a measure doesn't
do a good job when turnout is equal, then we see no reason to suppose
it will be useful in more taxing scenarios. There are also two
practical reasons for our choice. First, not all of the measures we
consider have been clearly defined for unequal turnout, although there
are reasonable extensions for each of them. Second, we feel that
considering multiple measures on multiple vote distributions with
multiple variations in turnout is unmanageable for a single
article. So, as much as unequal turnout is an important factor that
requires further investigation, we defer it to another article.

Our focus in this article is on moderate-sized elections in which both
parties enjoy significant statewide support. We have chosen an
arbitrary threshold of seven seats for our historical explorations in
Section~\ref{sec:hist}; our hypothetical elections in
Section~\ref{sec:hypo} all have at least ten seats and none are in the
hundreds, as is the case for many state legislatures. We set a minimum
size because we expect the strengths and weaknesses of the measures to
be significantly different in small elections. We note that the
declination is undefined when one party sweeps the election, as
happens, for example, 63\%, 32\% and 3\% of the time when there are
two, four and six seats, respectively. Only one percent of elections
with at least seven seats are swept by one party. Relatedly, the
measure $\noscale{\dg}$ won't consider any election as fair when one
party garners more than 66\% of the statewide vote as the dominant
party will necessarily waste votes. This happens frequently in small
states (for example, 9\% of elections with five seats), but occurs
only about one percent of the time when there are at least seven
seats.

Finally, we also limit our analysis to the realm of quantification of
gerrymandering: We do not include any discussion of the legal aspects
of the gerrymandering problem. Discussions such as those
in~\citep{McGann-elj} on the constitutional basis for supporting a
given measure are beyond the scope of this article. Nor do we attempt
to create a ``manageable standard'' that can be readily used by the
courts (see, e.g.,~\citep{M-S} or~\citep{Wang}).  Such aspirations,
while laudable, add a completely new dimension to the
issue. Furthermore, quantitative measures of gerrymandering can prove
useful outside of the courts. Even if a measure cannot directly serve
as justification for the invalidation of a district plan on
constitutional grounds, it can still serve as an aid to a
redistricting commission evaluating submitted plans, for educational
outreach, or for helping our understanding of issues such as the
effects of partisan gerrymandering (see, e.g.,~\cite{Warshaw}).

\subsection{Outline of article}

In Section~\ref{sec:semantics} we explore what it means to be a
partisan gerrymander. As part of this exploration, we present two
postulates describing a class of elections that should be evaluated as
fair and another that should be evaluated as unfair. We introduce the
measures we consider in this article in Section~\ref{sec:measures}. In
Section~\ref{sec:hist} we evaluate the measures on historical
elections. In addition to their inherent interest as data, the
distributions generated in this step are necessary for understanding
how to interpret individual values of each measure. In
Section~\ref{sec:hypo} we move on to evaluating hypothetical elections
in order to identify scenarios under which measures return false
positives or false negatives. Section~\ref{sec:sens} explores the
sensitivity of each measure to statewide vote share. We close in
Sections~\ref{sec:disc} and~\ref{sec:conc} with a discussion and our
conclusions, respectively.

\section{Gerrymandering, gerrymanders and partisan asymmetry}
\label{sec:semantics}

In order to evaluate gerrymandering measures, we need to describe
carefully what it is we wish to measure. We will distinguish in the
following way between the verb \emph{gerrymandering} and the noun
\emph{gerrymander}. For us, gerrymandering is the process of
attempting to create a map that illicitly favors one or more
groups. The extent to which the gerrymanderers are successful in
achieving their aims may have legal importance, but it does not affect
their intent. In this sense, determining whether or not someone is
gerrymandering is not, assuming full information, a difficult
question: What are the motives for the choices they are making?

Defining a gerrymander is a trickier matter. It can't \emph{just} be
the lines on a map --- if all the residents of a state are avowed
Democrats, then strange boundaries do not directly serve a partisan
purpose of the sort we are interested in. On the other hand, we can't
expect a measure such as those we will discuss in this article to be
able to know the intent of the map drawers. Since part of being a
partisan gerrymander is, according to our definition, the fact that it
has been drawn illicitly, no function based solely on the distribution
of votes can definitively identify a district plan as a
gerrymander. Our solution is to evaluate the proposed measures as
measures of partisan asymmetry: Are the two parties treated equitably
by the district plan? Any inequity that does occur may stem from
geography, be accidental, or as the result of intentional
gerrymandering; by itself, the measure can take no meaningful stand on
the source of the inequity.

The next step is to define what we mean by partisan asymmetry. This in
itself is a difficult matter as any precise definition is tantamount
to defining a measure of the type we are trying to independently
evaluate. We therefore take the indirect approach of articulating
identifiable characteristics of fair district plans and of unfair
ones. Once this is done, we will be in a position to label the various
elections from Section~\ref{sec:hypo} as either false positives or
false negatives for individual measures. We aim in the below to take a
minimal set of two postulates for identifying these characteristics. A
different choice of postulates would, of course, lead to different
conclusions regarding false positives/negatives.\\

\noindent
\textbf{Postulate 1:} A linear vote distribution should be
characterized as fair.\\

By ``linear'' we mean the Democratic vote fractions, when sorted in
increasing order, lie approximately along a line. All slopes are
considered equally appropriate. One notable property of these
distributions is that when a uniform vote shift is applied to give
each party equal statewide support, each party wins half of the
seats. It is also worth mentioning that these distributions yield
symmetric seats-votes curves. Finally, Postulate 1 is merely stating a
sufficient condition on what it means to be fair; it does not place
any restrictions on the other vote distributions that should also be
considered fair.

The slope of such a linear vote distribution is usually expressed in
terms of responsiveness or proportionality: $m$-proportionality occurs
when $(50 + x)\%$ of the statewide vote leads to $(50 + m\cdot x)\%$
of the seats. Historically, elections often show between $2$- and
$3$-proportionality (see~\citep{tufte}), but we intentionally do not
enshrine any particular proportionality constant as the one correct
value. The ``correct'' responsiveness, if there is one, is as
suggested by~\citep[Fig. 2C \& D]{declination}, dependent on how
partisans are distributed geographically. If the two groups are
distributed homogeneously, we would expect all districts to have
essentially the same vote split --- this leads to very high
responsiveness. Achieving low responsiveness, and the concomitant
uncompetitive districts, would likely require contorted districts so
as to string together local pockets where the majority party is
slightly more dominant than the statewide average. By restricting our
attention to vote distributions, we are foregoing important geographic
information that is crucial for illuminating what is or is not a
partisan gerrymander.  This is why it is better to view vote
distribution-based measures as identifying asymmetry or inequity,
rather than as a conclusive technique for ascribing an asymmetry to
partisan gerrymandering.\\

\noindent
\textbf{Postulate 2:} A vote distribution should be characterized as
unfair if the average winning margin for the party winning a majority
of the seats is significantly less than the average winning margin for
the opposition party.\\

Postulate 2 encodes the direct consequences of packing and cracking:
The minority party wastes votes in a few overwhelming wins while the
majority party efficiently wins a large number of relatively narrow
victories. As additional support for Postulate 2, we observe that the
2012 North Carolina US House election has a distribution exactly of
the form described. This district plan sets off alarm bells no matter
how one looks at it: the district plan was drawn by one
party~\citep{loyola:nc}; a politician involved in the process is on
record admitting that it was a partisan gerrymander~\citep{lewis}; the
seats won by the Democrats are far below their statewide vote (4 of 13
seats with 50.60\% of the total vote~\citep{wiki:nc12}); and the
proportion of seats won by Democrats is significantly less than what
they won in 2010 even though their statewide support improved over
2010 (Democrats won 7 of 13 seats in 2010 with 45.24\% of the total
vote~\citep{wiki:nc10}). A measure that doesn't mark the 2012 North
Carolina House election as an outlier is missing the closest thing we
have to an a universally acknowledged partisan gerrymander.

Absent from our two postulates is any mention of competitiveness.
This is because our primary goal in this article is to evaluate
measures of \emph{partisan} asymmetry. While lack of competitiveness
might be indicative of bipartisan (incumbency) gerrymandering, it has
no direct bearing on partisan gerrymandering. Said another way,
assessing a district plan according to whether it will aid or hinder
competitiveness is a worthy goal. But we consider competitiveness as a
confounding factor for our question and one that should not provide a
strong signal to partisan-asymmetry measures.

Nonetheless, the degree of competitiveness should provide \emph{some}
signal. It would be a stretch to call an election the result of a
gerrymander if the Democrats lose three districts with 49.3\% of the
vote while winning seven others with 50.3\% of the vote. But a much
stronger claim of gerrymandering could be levied if they lose three
districts with 36\% of the vote while winning seven others with 56\%
of the vote. Yet all that has changed is a factor-of-20 increase in
winning margins.

Another quality of measures seen as desirable is stability under
changing electoral climates. In fact, some authors\footnote{For
  example: ``Moreover, if a standard sometimes identifies the same set
  of districts as a gerrymander with respect to some elections and a
  non-gerrymander with respect to other elections, we know with
  assurance it is committing errors,'' ~\citep[pg. 4]{McDonald-two}.}
argue that a gerrymander exists independently of the ebb and flow of
partisan support and that measures should recognize the gerrymander
for what it is regardless of how the statewide support for each party
changes. We do not believe this is most efficacious point of view. If
one party suddenly becomes wildly popular, it likely doesn't matter at
all how the lines are drawn: the opposing party isn't going to win a
single seat. A simple function, based only on the partisan vote
distribution, isn't going to be able to recognize gerrymandering that
expected a completely different electoral climate. Less drastically,
it seems reasonable, even preferable, that a measure return a maximum
value of unfairness for a given level of partisan support and
gradually reduce in value as support deviates from the expectation.

We view too much stability as a sign that the measure is not capturing
the essence of a partisan gerrymander: the ability to turn votes into
seats. On the other hand, if a measure is \emph{too} sensitive to such
swings then it becomes useless. A measure that frequently
registers a strong pro-Democratic advantage with a given statewide
level of Democratic party support but a strong pro-Republican
advantage under a uniform partisan swing of one percent is not going
to be very useful.

Together these issues lead us to the following principle to use as a
secondary basis on which to evaluate our measures. We state this
principle in order to explicitly acknowledge that there are other
important grounds on which to evaluate measures. These other grounds
are related to, but distinct from, false positives and false
negatives. We do not attempt to specify precisely how much ``too
much'' or ``too little'' is.

\noindent
\textbf{The Goldilocks Principle:} A measure should have neither too
much nor too little sensitivity to both competitiveness and overall
statewide support.

\section{Gerrymandering measures}
\label{sec:measures}

In this article we will view an \emph{election with $N$ districts} as
a triple consisting of two parties $D$ (Democrats) and $R$
(Republicans) along with a sequence
\begin{equation}\label{eq:bp}
  \bp = 0\leq p_1 \leq p_2 \leq \cdots \leq p_k \leq \frac{1}{2}
  < p_{k+1} \leq p_{k+2}\leq \cdots \leq p_N\leq 1,
\end{equation}
where $p_i$ denotes the fraction of the two-party vote won by Party
$D$ in district $i$. We display our elections by
plotting the $p_i$ values along a vertical axis with districts indexed
along the horizontal axis. Let $\barp$ be the statewide fraction of
the vote attained by Party $D$.

We have chosen to focus in this article only on those measures that
depend solely on $\bp$. It is worth mentioning that there are other
non-geographic measures that are excluded by this criterion. For
example, in~\citep{Mattingly17}, the \emph{gerrymandering index} and
the \emph{representativeness index} are defined as measures of
gerrymandering. However, these measures utilize simulated district
plans in their definitions in an integral manner. Fortunately, all of
the measures we do consider in this article can be combined with
computer simulations if so desired.

The measures and acronyms defined in this section are summarized in
Table~\ref{tab:measures}.

\subsection{The efficiency gap and its variants}

The first class of measures we consider all rely on the notion of a
\emph{wasted vote}. We should emphasize that all of the measures trace
their lineage back to the efficiency gap (which we will write as
$\noscale{\eg}$) introduced in~\citep{McGhee} and explored
in~\citep{M-S}. Each of the variants introduced was an attempt to
improve upon it.

In the context of the original efficiency gap, a wasted vote is a vote
for a losing candidate or a vote for a winning candidate that is in
excess of the 50\%+1 needed to win. Following others
(see~\citep{Cover,Nagle1,Nagle2}) we define a slightly more general
function that incorporates a parameter $\lambda$:
\begin{equation*}
  \waste_D^\lambda(i) =
  \begin{cases}
    p_i, & \text{if Party $D$ loses district $i$},\\
    \lambda\cdot (p_i-1/2), & \text{if Party $D$ wins district $i$}.
  \end{cases}
\end{equation*}
Note that as defined, $\waste_D^\lambda(i)$ actually keeps track of
the \emph{fraction} of votes in district $i$ that are wasted rather
than the total number of votes. For each district we define
$\waste_R^\lambda(i)$ analogously. The efficiency gap and its variants
all work by comparing the number of votes wasted by Party
$D$ to the number wasted by Party $R$. For general $\lambda$, we can
define (after~\citep{Nagle2}) the \emph{weighted efficiency gap},
\begin{equation}\label{eq:lambdagap}
    \noscale{\lambdaeg{\lambda}} = \frac{\sum_{i=1}^N (\waste_D^\lambda(i)-\waste_R^\lambda(i))}{N}.
\end{equation}

The only values of $\lambda$ we know of that have been seriously
considered in the literature are $\lambda \in \{0,1,2\}$, so we
likewise restrict our attention to these. The \emph{efficiency
  gap}~\citep{McGhee,M-S} is $\noscale{\eg}$. It has been suggested in
several places~\citep{Griesbach,Cover,Nagle2} that wasted winning votes
in a district should be relative to the losing party vote rather than
to the 50\% threshold. This can be achieved by considering
$\noscale{\dg}$. Also considered in~\citep{Nagle2} is the case in which
only losing votes are counted as wasted. This leads to the use of
$\noscale{\lossg}$. We note that $\noscale{\eg}$ will not find
\emph{any} election fair if one party garners more than 75\%; the
same is true for $\noscale{\dg}$ at a threshold of 66\%.

In~\citep{cho-upenn} the author asserts that winning and losing wasted
votes should not be treated equally. She proposes a \emph{winning
  efficiency} that does not include votes wasted towards losses at
all:
    \[ \frac{\sum_{i=k+1}^N \waste_D^1(i) - \sum_{i=1}^k \waste_R^1(i)}{2N}.\]
It turns out that the winning efficiency reduces to the difference in
statewide support between the two parties. As such, it considers an
election to be fair if and only if the parties each garner 50\% of the
statewide vote. Statewide support is an environmental factor that is
not, by itself, indicative of whether or not a gerrymander has been
drawn. This makes it useless for quantifying partisan gerrymanders and
we discuss it no more in this article.

The authors in~\citep{Nagle2} and~\citep{Cover} suggest comparing the
fraction of Democratic votes that are wasted to the fraction of
Republican votes that are wasted. Given a parameter $\lambda$ we
define
    \begin{equation*}
      \noscale{\vcg^\lambda} = \frac{\sum_{i=1}^N \waste_D^\lambda(i)}{\sum_{i=1}^N p_i} - 
      \frac{\sum_{i=1}^N \waste_R^\lambda(i)}{\sum_{i=1}^N(1-p_i)}.
    \end{equation*}
We consider here only the two ``vote-centric'' versions of the
efficiency gap given by $\noscale{\va}$ and $\noscale{\vb}$.

Finally,~\citep{declination} introduces the
\emph{$\tau$-Gap} measure which weights votes according to a function
parameterized by a non-negative real number $\tau$. The idea behind the
definition is that votes close to the fifty-percent threshold are less
wasted than those at either extreme. When $\tau=0$, the $\tau$-Gap
reduces to twice the efficiency gap. The precise definition for
general $\tau$ requires an integral which we do not attempt to
replicate here. We will restrict our attention in this article to the
value of $\tau = 1$ and denote the measure by $\noscale{\taugap}$.

\subsection{The declination and a variant}
\label{sec:dec}

The declination, introduced in~\citep{declination} (see
also~\citep{DecIntro}), is essentially an angle associated to the vote
distribution. It can be thought of as a measure of differential
responsiveness, albeit one that is anchored at the actual statewide
level of support between the two parties rather than at 50\%. It
depends on four quantities: the average winning margin of each party
and the fraction of seats each party wins.

Recall from equation~\eqref{eq:bp} that there are assumed to be $k$
districts won by Party $R$ out of $N$ total. Let $k'$ denote the
number of districts won by Party $D$ (so $k+k'=N$).  Define half of
the average winning margin for parties $R$ and $D$, respectively, as
\begin{equation}\label{eq:yz}
  y = \frac{1}{k}\sum_{i=1}^k (1/2-p_i)\text{ and }
  z = \frac{1}{k'}\sum_{i=k+1}^N (p_i-1/2).
\end{equation}

\ifdef{\SUBMIT}{}{
\begin{figure}
  \centering
  \includegraphics[width=.6\linewidth]{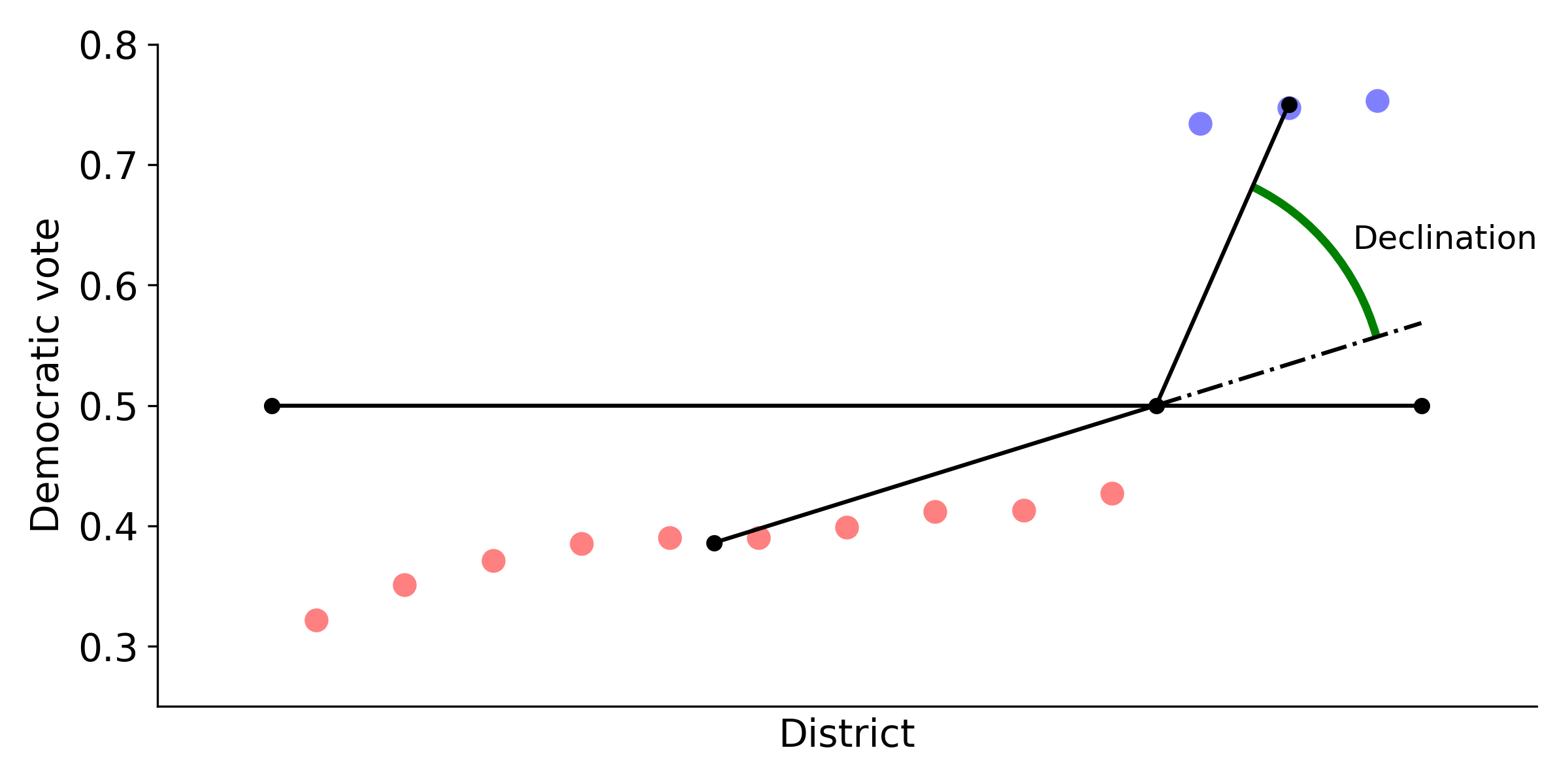}
  \caption{\mycapb}
  \label{fig:plot}
\end{figure}
}

If $k=0$ or $k'=0$, the declination is undefined. Otherwise, we set
$\theta_D = \arctan\left(\frac{2z}{k'/N}\right)$ and $\theta_R =
\arctan\left(\frac{2y}{k/N}\right)$ and define the declination 
as $\noscale{\dec}=2(\theta_D-\theta_R)/\pi$ (see Figure~\ref{fig:plot}). Note that the declination is not
using the Democratic vote in the median district won by Democrats, but
rather the mean of these values.

The \emph{buffered declination}, $\noscale{\bdec}$, is computed
analogously to the declination except that we introduce $m=\lceil
N/20\rceil$ districts with $p_i = 0.5$ and $m$ districts with $p_i =
0.5001$ into the election (so $k$ and $k'$ are each increased by $m$
in equation~\eqref{eq:yz}). The modifications of $y$ and $z$ can be
interpreted as adding $m$ evenly split districts to those the
Democrats win and $m$ evenly split districts to those the Republicans
win. The rationale for doing this is that it buffers the angle in
cases for which one side wins a small fraction of the seats; this
reduces the sensitivity to the exact vote fractions in the minority
seats. When the sides are evenly matched, there is little effect
besides an overall reduction in values.

\subsection{Measures arising from the seats-votes curve}
\label{sec:misc}

The \emph{mean-median difference}, defined as $\noscale{\mm} = \barp -
\mathrm{median}(\bp)$, has been considered by a number of authors such
as in~\citep{Nagle1, Wang}. Here $\barp$ is the average of the $p_i$, or,
assuming equal turnout in all districts, the statewide vote fraction
for Party $D$. For $N$ odd, the median is $p_{(N-1)/2}$; for $N$ even,
the median is defined as the average of $p_{N/2}$ and $p_{1+N/2}$.

\emph{Partisan bias}~\citep{GK} aims to evaluate the degree to
which the two parties would win different fractions of the total seats
for the same level of popular support. We consider two versions. The
first compares the seat share each party would win if support were to
be uniformly shifted to 50\% for each party:
\begin{equation*}
  \noscale{\bias} = \frac{1}{2} - \frac{|\{i:\, p_i > \barp\}|}{N}.
\end{equation*}
For the second share, we compare the seat share the Democrats hold at
the level of $\barp$ for statewide support with the seat share the
Republicans would hold at the level of $1-\barp$ statewide
support. The partisan bias at the observed level of support is then:
\begin{equation*}
  \noscale{\biasO} = \frac{|\{i:\, p_i + 1 - 2\barp < 1/2\}| - |\{i:\, p_i > 1/2\}|}{2N}.
\end{equation*}
This variant is referred to as the \emph{specific asymmetry}
in~\citep{mcauliffe}.

\subsection{Conditioned measures}

The measures of the previous three subsections provide quantitative
measures, when defined, for any election. The following two proposals
are partially tests, in the sense of~\citep{McGheeTest}, rather
than pure measures: The focus is not only on the numerical value
associated to the election, but also on whether a given condition is
satisfied and, hence, whether a threshold has been crossed that
warrants remediation.

The \emph{lopsided means} measure, $\noscale{\lm}$, proposed
in~\citep{Wang}, computes the average winning margin of Party $D$
minus the average winning margin of Party $R$. In the notation
of~\eqref{eq:yz}, this is just $z-y$. While this can be used as its
own, standalone measure, Wang argues for using this measure in
conjunction with a Student $t$-test. Under this framework, the measure
only indicates gerrymandering if the difference in margins is
statistically significant at $p < 0.05$. For most of this article we
will work with the measure alone, however in the discussion we explore
some of the issues involved with invoking a statistical test in this
context.

The \emph{equal vote weight} standard~\citep{McDonald-two},
$\noscale{\evw}$, is equivalent to the mean-median measure, but
subject to the constraint that a gerrymander is indicated exactly when
the party winning a majority of the statewide vote wins a minority of
the seats, i.e., an anti-majoritarian outcome. Because of the
numerical equivalency with the mean-median measure, $\noscale{\evw}$
will not figure prominently in this article, however we will return to
it in the discussion.

\ifdef{\SUBMIT}{}{
\small
\begin{table}
\centering
\caption{Summary of measures considered and their acronyms.}
\begin{tabular}{lp{5in}}
\toprule
Acronym & Measure\\
\midrule
$\noscale{\eg}$ & \emph{Efficiency gap:} comparison of wasted votes.\\
$\noscale{\dg}$ & \emph{Margin efficiency gap:} Variant of $\noscale{\eg}$ in which wasted votes are relative to losing party vote.\\
$\noscale{\lossg}$ & \emph{Losing efficiency gap:} Variant of $\noscale{\eg}$ in which only losing wasted votes are compared.\\
$\noscale{\va}$ & \emph{Vote-centric efficiency gap:} Variant of $\noscale{\eg}$ in which each comparison is between the fraction of votes wasted by each party rather than between the absolute amounts.\\
$\noscale{\vb}$ & \emph{Vote-centric losing efficiency gap:} Combination of $\noscale{\dg}$ and $\noscale{\va}$.\\
$\noscale{\taugap}$ & \emph{$\tau$-Gap:} Variant of $\noscale{\eg}$ in which votes are weighted according to a function parameterized by $\tau$.\\
$\noscale{\dec}$ & \emph{Declination:} A measure of differential responsiveness.\\
$\noscale{\bdec}$ & \emph{Buffered declination:} Variant of $\noscale{\dec}$ that is less sensitive in cases in which one party wins most of the votes.\\
$\noscale{\mm}$ & \emph{Mean-median difference:} A comparison of the median and mean of distribution.\\
$\noscale{\bias}$ & \emph{Partisan bias:} A comparison of the seat share at 50\% of the statewide vote.\\
$\noscale{\biasO}$ & \emph{Specific asymmetry:} Variant of partisan bias that focuses on symmetry relative to observed vote.\\
$\noscale{\lm}$ &  \emph{Lopsided means:} A comparison of the average winning margins of parties.\\
$\noscale{\evw}$ & \emph{Equal vote weight:} Variant of $\noscale{\mm}$ that requires an anti-majoritarian outcome for a positive result.\\
\bottomrule
\label{tab:measures}
\end{tabular}
\end{table}
}

\section{The measures on historical data}
\label{sec:hist}

In Section~\ref{sec:hypo} we will evaluate a number of hypothetical
elections using the measures introduced above. However, the resulting
values will be difficult to interpret without some understanding of
the distribution of values we should expect to arise for each
measure. (Does $\noscale{\va} = 0.1$ indicate an outlier for this
accounting of wasted votes, or an election well within historical
norms?) To this end, we first evaluate the measures on a large set of
historical elections. With these data we can calculate a standard
deviation for each measure. These deviations allow us to compute the
extent to which each hypothetical is marked, or is not marked, as an
outlier.

Our historical data set consists of state-level lower house elections
since 1972 for which there are no multi-member districts, as well as
US House elections since 1972. This data set, used in~\citep{redact},
is described in detail therein. Two points are worth mentioning
here. First, we ignore third-party candidates: all $p_i$ represent
fractions of the two-party vote\footnote{Our data for congressional
  races do not include third-party candidates, but we note for
  reference that about 16\% of state legislative races in our data
  include a third-party candidate who won at least one percent of the
  total vote.}. Second, in cases in which there are not candidates for
both parties, we impute the vote fraction\footnote{Races that are not
  contested by both parties occur in some states two-thirds of the
  time for state-level house elections --- Wyoming, Georgia, Arkansas
  and South Carolina all slightly exceed this rate. There is no
  obvious correspondence between high rates on uncontestedness and
  gerrymandering, although we have not delved into the data in more
  than a superficial manner. Rates for congressional elections are
  much lower, the only states exceeding 30\% are Alabama, Louisiana
  and Massachusetts.}. Both issues are important for any application
of these measures to real-world elections. In this article, however,
we are focused on individual measures rather than identifying the
particular states and years for which they find unfairness, so we do
not dwell on these issues.

The partisan vote we assign to each district can come from either an
actual election the district was used for or from exogenous election
data such as a statewide gubernatorial race. Each approach has its
staunch advocates (see, e.g., the conversation
in~\citep{McGheeRejoinder} and~\citep{BestResponse}). In this article we
use the endogenous, two-party vote share. Using exogenous data would
certainly change the particular elections that appear as extreme
examples in Table~\ref{tab:outliers}, in the appendix, and the exact
values in Table~\ref{tab:hypo}.

\ifdef{\SUBMIT}{}{
\small
\begin{table}
\centering
\caption{Mean and standard deviation for the measures considered in
  this article applied to the 1,166 elections in our data set with at
  least seven seats for which each party wins at least one seat.}
\begin{tabular}{lrrrrrrrrrrrrrr}
\toprule
{} &   $\noscale{\eg}$ &   $\noscale{\dg}$ &    $\noscale{\lossg}$ &   
   $\noscale{\va}$ &   $\noscale{\vb}$ & $\noscale{\taugap}$ 
&  $\noscale{\dec}$ &  $\noscale{\bdec}$ &    $\noscale{\mm}$ &
$\noscale{\bias}$ &  $\noscale{\biasO}$ & $\noscale{\lm}$ &   $\noscale{\evw}$\\
\midrule
Mean    & 0.01 &             0.04 &            -0.02 &              -0.05 &              -0.01 & -0.04 & -0.01 &  0.01 & -0.00 & -0.01 &               0.01 & 0.03 & -0.00 \\
Std Dev & 0.07 &             0.08 &             0.11 &               0.23 &               0.17 &  0.25 &  0.20 &  0.11 &  0.04 &  0.09 &               0.06 & 0.06 &  0.04 \\
\bottomrule
\label{tab:mean}
\end{tabular}
\end{table}
}

In Table~\ref{tab:mean}, we record the mean and standard deviation for
each measure on the collection of historical elections. Positive
values indicate Republican advantages while negative values indicate
Democratic advantages. In Section~\ref{sec:hypo} we will be comparing
how different hypothetical elections are evaluated by different
measures.  For this purpose it will be convenient for the remainder of
the article to rescale each measure so it has standard deviation equal
to 1. This is done by dividing each of the formulas in
Section~\ref{sec:measures} by the corresponding standard deviation
from Table~\ref{tab:mean}.

In Figure~\ref{fig:risk-kde} we display kernel density plots for each
rescaled measure on the historical elections in our data set. The plot
has been split into three to avoid clutter; there is no significance
to which of the three subplots a given rescaled measure
appears. Note that with these unitless rescaled measures, a value of
(say) $2$ denotes two standard deviations above the mean. Also note
that we are \emph{not} normalizing the measures --- that is we are not
shifting the mean to zero.

Even though most of the measures we discuss generate ``approximately
normal'' distributions from our historical corpus in
Figure~\ref{fig:risk-kde}, the standardization we apply does not
depend on an assumption of normality. It is perfectly
meaningful to discuss standard deviations for non-normal
distributions. 

\ifdef{\SUBMIT}{}{
\begin{figure}
  \centering
  \includegraphics[width=1\linewidth]{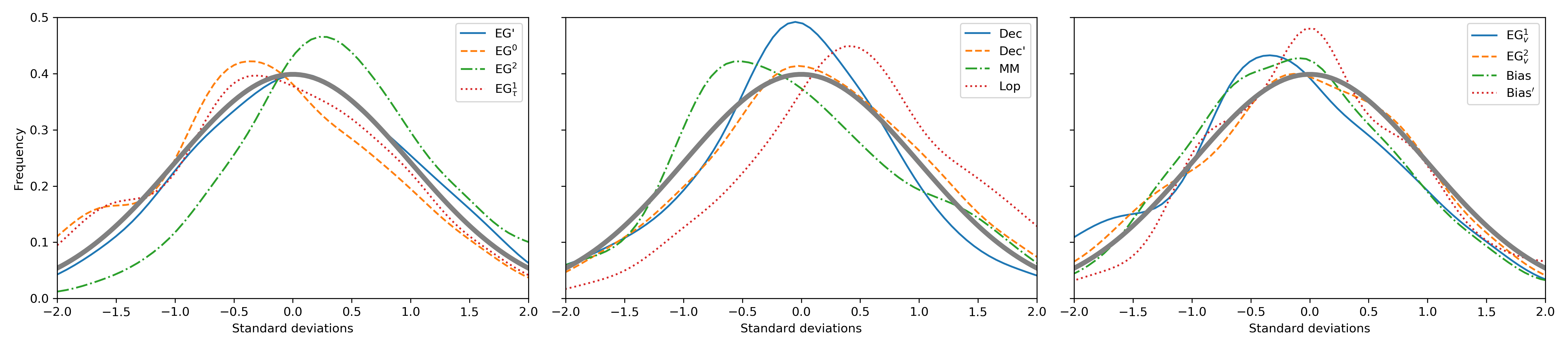}
  \caption{\mycapc}
  \label{fig:risk-kde}
\end{figure}
}

The next two subsections explore the aggregate data in more
depth.

\subsection{Correlations among measures}
\label{sec:corr}

Before exploring how the measures described in
Section~\ref{sec:measures} work on individual elections, it is first
worth considering how they correlate in the aggregate
(cf.~\citep{MSII}). We show in Figure~\ref{fig:measure-scatter} scatter
plots of various pairs of measures. Note that $\scale{\evw}$ has been
omitted since it is numerically equivalent to $\scale{\mm}$.
As $\scale{\dec}$ is not defined when one party sweeps all seats, the
thirteen such races have been omitted for consistency from all
comparisons in the figure. Each scatter plot thereby consists of 1,166
points. 
The ``bow tie'' appearance of the scatter plot for $\scale{\mm}$ and
$\scale{\bias}$ arises from the fact, noted in~\citep{MSII} and
perhaps elsewhere, that $\scale{\mm}$ yields $\scale{\bias}$ when
multiplied by the responsiveness.

\ifdef{\SUBMIT}{}{
\begin{figure}
  \centering
  \includegraphics[width=1\linewidth]{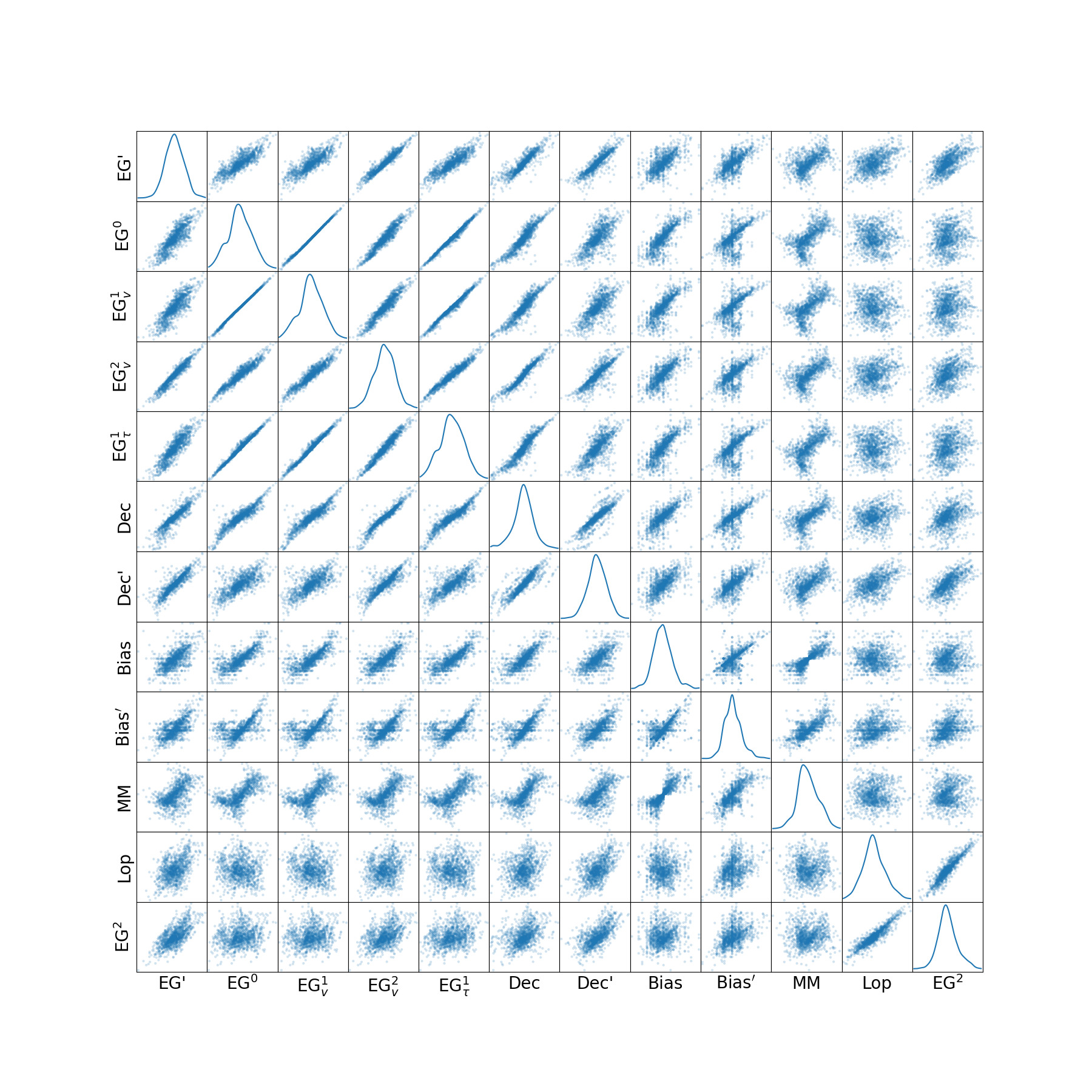}
  \caption{\mycapd}
  \label{fig:measure-scatter}
\end{figure}
}

\subsection{Historical outliers}
\label{sec:evaluations}

\ifdef{\SUBMIT}{}{
{\tiny
  \begin{table}
    \centering
    \caption{Outliers listed as last two digit of year along with
      state abbreviation. The first group of ten corresponds to the
      population of the 647 most-balanced elections. The second
      group of ten corresponds to the population of all 1,179
      elections. In each grouping, the most extreme outliers are
      listed first. An asterisk denotes a state, rather than US House,
      election.}
\begin{tabular}{@{} cllllllllllllll @{}}
\toprule
& {} &       EG & $\mathrm{EG}^2$ & $\mathrm{EG}^0$ & $\mathrm{EG}_v^1$ & $\mathrm{EG}_v^2$ &  $\mathrm{EG}_\tau^1$ & Dec &    Dec' &      MM &    Bias & Bias' &     Lop &     EVW \\
\cmidrule{2-14}
& 1  &  78 WA  &          78 WA  &          78 WA  &            78 WA  &           78 WA  &  78 WA  &  78 WA  &  12 PA  &  00 TN  &  96 WA  &            14 NC  &  96 WA  &  80 IL  \\
& 2  &  94 WA  &          94 WA  &          94 WA  &            94 WA  &           94 WA  &  12 OH  &  12 OH  &  12 OH  &  06 GA  &  16 NC  &            04 AZ  &  80 WA  &  04 MI  \\
& 3  &  12 OH  &          06 MI  &          12 OH  &            12 OH  &           12 OH  &  12 PA  &  14 NC  &  16 PA  &  06 TN  &  12 NC  &            12 OH  &  94 WA  &  88 KY  \\
& 4  &  12 PA  &          96 WA  &          14 NC  &            12 IN  &           12 PA  &  14 NC  &  12 PA  &  14 NC  &  06 MO  &  14 NC  &            12 NC  &  94 AL  &  88 WI  \\
& 5  &  94 MN  &          12 PA  &          12 IN  &            14 NC  &           94 MN  &  94 WA  &  94 WA  &  12 MI  &  80 IL  &  12 OH  &            12 VA  &  98 AL  &  98 TX  \\
& 6  &  96 WA  &          12 MI  &          94 MN  &            94 MN  &           14 NC  &  16 NC  &  12 VA  &  06 MI  &  80 KY  &  04 AZ  &            12 PA  &  12 WI  &  84 WI  \\
& 7  &  12 NC  &          10 IL  &          16 NC  &            16 NC  &           12 VA  &  12 IN  &  16 NC  &  14 PA  &  06 AL  &  06 VA  &            96 WA  &  14 NJ  &  78 IL  \\
& 8  &  12 VA  &          12 OH  &          12 PA  &            12 PA  &           96 WA  &  16 PA  &  16 PA  &  92 TX  &  96 VA  &  12 VA  &            16 PA  &  94 VA  &  02 IL  \\ \rot{\rlap{Balanced elections}}
& 9  &  14 NC  &          12 NC  &          12 VA  &            12 VA  &           12 NC  &  94 MN  &  72 KY  &  12 NC  &  08 GA  &  12 PA  &            80 WA  &  84 TN  &  14 MI  \\
& 10  &  80 WA  &          06 OH  &         06 VA &            06 VA  &            12 IN  &  12 VA  &  06 VA  &  10 IL  &  04 MO  &  16 PA  &            72 KY  &  72 MN  &  16 MI  \\ \cmidrule{2-14}
& 1 &  10 MA  &          06 NY* &          10 MA  &            10 MA  &            10 MA  &  10 MA  &  80 VA  &  92 HI* &  74 AL  &  16 SC  &            14 NC  &  86 MD  &  80 IL  \\
& 2 &  78 WA  &          08 NY* &          14 MA  &            14 MA  &            78 WA  &  14 MA  &  90 MA  &  12 PA  &  72 AL  &  14 AL  &            04 AZ  &  82 AL  &  04 MI  \\
& 3 &  94 WA  &          06 NY  &          02 MA  &            02 MA  &            14 MA  &  02 MA  &  76 TX  &  12 OH  &  00 TN  &  12 AL  &            12 OH  &  74 MD  &  88 KY  \\
& 4 &  92 WA  &          00 NY* &          12 MA  &            12 MA  &            80 VA  &  80 VA  &  79 MS* &  78 TX* &  06 GA  &  86 MA  &            12 NC  &  92 MA  &  88 WI  \\
& 5 &  80 VA  &          16 NY  &          78 WA  &            16 MA  &            92 WA  &  12 MA  &  92 HI* &  76 TX  &  06 TN  &  92 WA  &            12 VA  &  98 MD  &  98 TX  \\
& 6 &  12 OH  &          96 NY* &          80 VA  &            04 MA  &            94 WA  &  78 WA  &  72 GA  &  94 HI* &  80 AL  &  14 IN  &            12 PA  &  74 NC  &  84 WI  \\
& 7 &  14 MA  &          98 NY* &          92 WA  &            76 GA  &            12 SC  &  79 MS* &  88 MA  &  16 PA  &  06 MO  &  96 WA  &            96 WA  &  96 MD  &  78 IL  \\ \rot{\rlap{All elections}}
& 8 &  12 PA  &          72 LA* &          79 MS* &            74 GA  &            02 MA  &  04 MA  &  12 SC  &  86 AL* &  86 WA  &  16 TN  &            16 PA  &  96 WA  &  02 IL  \\
& 9 &  12 SC  &          04 NY* &          16 MA  &            98 MA  &            12 MA  &  72 MO  &  84 MA  &  14 NC  &  86 FL  &  16 NC  &            72 KY  &  74 MN  &  14 MI  \\
& 10 &  94 MN  &          78 WA  &         04 MA  &            08 MA  &            12 OH  &  12 SC  &  86 MA  &  76 TX* &  80 IL  &  14 NC  &            98 AL  &  72 WI  &  16 MI  \\
\bottomrule
\end{tabular}
\label{tab:outliers}
\end{table}
}
}

We close our summary of historical elections by recording in
Table~\ref{tab:outliers} the outliers for each measure among two
different populations. This material is not used in the rest of the
article and is provided for the reader who is interested in seeing
what each measure determines to be the most extreme historical
outliers.

For the first population, we restrict our attention to elections with
at least seven seats in which each party holds statewide support
between 45\% and 55\%. There are 647 of these ``balanced'' elections
in our data set.

The second population considered is that of all elections with at
least seven seats. There are 1,179 such elections.  In elections that
are dominated by one party, the measures begin to break down. For
example, in an election in which one party receives more than two
thirds of the vote: the $3$-proportionality of $\scale{\dg}$ implies
that it is automatically unfair as one party cannot win more than
100\% of the seats; the declination will be undefined if one party
sweeps all of the seats; and the counterfactuals required for the two
versions of partisan bias become daunting. Fortunately, even if a
measure does poorly for imbalanced elections, it may still have a role
to play for balanced ones.

For the two declination variants and the lopsided-means test,
elections in which the measures are undefined are omitted from
Table~\ref{tab:outliers}. Similarly, for $\scale{\evw}$ only
anti-majoritarian elections are considered; for $\scale{\lm}$ only
elections in which the $t$-test indicates statistical significance are
considered.

\section{Hypothetical elections}
\label{sec:hypo}

In this section we introduce a number of hypothetical elections and
evaluate our measures on them.  By choosing a small set of
hypothetical elections of various types, we are able to identify
potential false positives and false negatives for each measure. These
false positives/negatives are, of course, relative to the framework of
Section~\ref{sec:semantics}. This classification of
fairness/unfairness might differ from the one made for an actual
historical election once additional factors, such as the geographic
distribution of partisans, are taken into account.

\ifdef{\SUBMIT}{}{
\begin{figure}
  \centering
  \includegraphics[width=.9\linewidth]{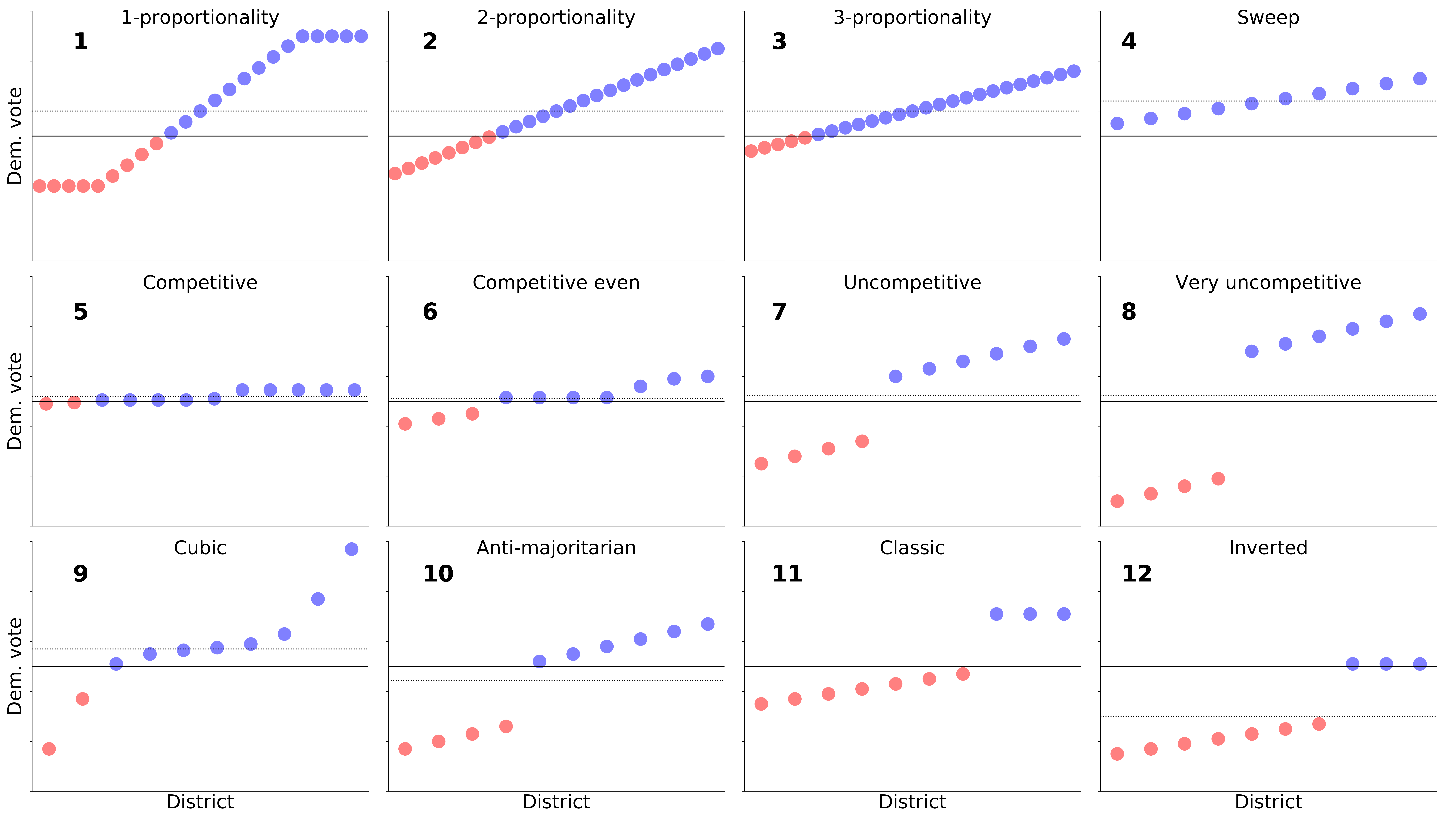}
  \caption{\mycapm}
  \label{fig:hypo_grid}
\end{figure}
}

\subsection{The twelve hypothetical elections}
\label{sec:hypo-defs}

Below are brief descriptions of the twelve hypothetical elections
considered in Figure~\ref{fig:hypo_grid}; evaluations of the measures
on these elections are provided in Table~\ref{tab:hypo}.

Our first four hypothetical elections should, according to Postulate
1, be evaluated as fair by a vote-distribution measure.

\begin{enumerate}
  \item[A] \textbf{$1$-proportionality} (\elecA). For moderate uniform swings,
    this election is the ideal vote distribution according to
    $\scale{\lossg}$ and $\scale{\va}$. 

  \item[B] \textbf{$2$-proportionality} (\elecB). The ideal vote distribution
    according to $\scale{\eg}$.

  \item[C] \textbf{$3$-proportionality} (\elecC). The ideal vote distribution
    according to $\scale{\dg}$.

  \item[D] \textbf{Sweep} (\elecD). Party $D$ has swept the election with a
    statewide vote share of 64\%. This particular election corresponds
    to a constant of proportionality equal to $5$. This election is
    similar to many recent Massachusetts House elections.
    (see, e.g., Figure~\ref{fig:extremal-un-dg} in the appendix).
\end{enumerate}

How the next four hypothetical elections should be evaluated is less
straightforward. None is precisely linear, though the first two are
very close to linear. These first two have been included due to the
presence of a number of very competitive districts. Elections \elecG\ and \elecH,
however, are notably uncompetitive. Neither of our two postulates is
particularly pertinent for these latter two, although a measure should
evaluate both similarly if it is to have low sensitivity to
competitiveness.

\begin{enumerate}
  \item[E] \textbf{Competitive} (\elecE). This election has a number of very
    competitive races that all have swung towards Party $D$. 

  \item[F] \textbf{Competitive even} (\elecF). Similar to the previous
    election, but with no districts in the ``counterfactual window''
    (i.e., between the majority party's statewide support and 50\% ---
    see~\citep{McGhee}).

  \item[G] \textbf{Uncompetitive} (\elecG). An uncompetitive election as might
    arise from a bipartisan gerrymander; average winning margins for
    Republicans and Democrats are 41 and 35 points, respectively.

  \item[H] \textbf{Very uncompetitive} (\elecH). The previous election with
    average Republican and Democratic margins of victory increased by
    30 and 20 points, respectively.
\end{enumerate}

The next three elections are squarely covered by Postulate 2 and
should be marked as unfair. Election \elecL, a uniform shift of
Election \elecK, is more ambiguous. As it is not linear, we cannot apply
Postulate 1 to conclude that it is fair. On the other hand, since the
average winning margin for the Democrats is smaller than the average
winning margin for the Republicans, we cannot apply Postulate 2 to
conclude it is unfair. In fact, the Democrats' margin is so much
smaller that it suggests, if anything, a Democratic advantage. If this
really were a Republican advantage, then in any reasonable electoral
climate, it would make much more sense to accept slightly less
comfortable wins for a chance to win three additional seats.

\begin{enumerate}
  \item[I] \textbf{Cubic} (\elecI). This is a symmetric distribution
    shifted moderately to the Democrats' favor.

  \item[J] \textbf{Anti-majoritarian} (\elecJ). This is election G subjected to
    a uniform swing in favor of the Republicans.

  \item[K] \textbf{Classic} (\elecK). This distribution has all of the
    hallmarks of a partisan gerrymander: While the parties are evenly
    split statewide, the Republicans win a significant majority
    through having a number of narrow victories in contrast to their
    Democratic opponents whose few victories are overwhelming. This
    distribution is very similar to the 2014 North Carolina US House
    election.

  \item[L] \textbf{Inverted} (\elecL). This is dual to \elecK. While the
    Republicans win a significant majority of the vote, the Democrats
    are still able to win a significant portion of the seats through
    very narrow victories.
\end{enumerate}

\subsection{Analysis of evaluations}
\label{sec:analysis}

In Table~\ref{tab:hypo} we record the evaluations of the measures on
the hypothetical elections introduced in the previous section. The
equal vote weights measure is not given its own column since it is
numerically equivalent to $\scale{\mm}$; the only anti-majoritarian
election is I. We note that the $t$-test for $\scale{\lm}$ was not
statistically significant for any of the elections.

In Section~\ref{sec:disc} we discuss the performance of each measure
in turn. For now we make only a few high-level comments organized
according to election. First we recall that the first six elections
(\elecA, \elecB, \elecC, \elecD, \elecE\ and \elecF) should be
evaluated as fair by Postulate 1; Elections \elecI, \elecJ\ and \elecK\
should be evaluated as unfair by Postulate 2; Elections \elecG\ and
\elecH\ are at most modestly unfair by Postulate 2; the situation of
Election \elecL\ is unaddressed by either postulate. In the table we
have placed in bold those entries corresponding to false
positives/negatives, as determined by a somewhat arbitrary threshold
of two standard deviations from the mean. As such, the bold entries in
the first six rows correspond to false positives. With two exceptions,
these are confined to the variants of the efficiency gap. First, the
lopsided-means standard, $\noscale{\lm}$, evaluates Election \elecA\ as
strongly in favor of the Republicans. Second, the two variants of
partisan bias come to very different conclusions regarding Elections
\elecE\ and \elecF.

Similarly, the bold entries for Elections \elecI, \elecJ\ and \elecK\
correspond to false negatives. For Elections \elecG\ and \elecH\ we note
that $\scale{\mm}$ does not adhere to the Goldilocks Principle as the
doubling of winning margins leads to a doubling of the mean-median
difference, although it is true that it treats both as
outliers. Election \elecL\ is, as mentioned, rather unusual. The
measures come to very different conclusions ranging from seeing it as
strongly favoring the Democrats to strongly favoring the
Republicans. As discussed above, we believe it is correct to conclude
that the election strongly favors the Democrats. The partisan bias
measure $\scale{\bias}$ comes to the surprising conclusion that it is
strongly in favor to the Republicans. The mean-median difference comes
to the same conclusion, though less strongly, while $\scale{\lossg}$,
$\scale{\va}$ and $\scale{\taugap}$ find the election to be fair.

We make one final note regarding the efficiency gap as it pertains to
these data. As part of~\citep{M-S}, the authors propose a possible
legal standard for identifying partisan gerrymandering incorporating a
threshold for the efficiency gap. They suggest a threshold of 0.08 for
state house plans and of two seats for congressional plans. One can
recover the unscaled efficiency gap scores by multiplying the values
in the $\noscale{\eg}$ column of Table~\ref{tab:hypo} by the standard
deviation of $0.07$. Using either threshold, one can then check that
the only three elections that do not surpass the threshold are \elecB, \elecG\
and \elecH.

\ifdef{\SUBMIT}{}{
\begin{table}
  \centering
\caption{Values of scaled measures on hypothetical elections from
  Figure~\ref{fig:hypo_grid}. Column $N$ provides the number of
  districts in each election while the statewide Democratic vote share
  is indicated in the Mean column. Bold entries indicate false
  positives/negatives.}
\begin{tabular}{lllllllllllllll}
\toprule
{} &   N &  Mean &             $\scale{\eg}$ &             $\scale{\dg}$ &             $\scale{\lossg}$ &     $\scale{\va}$ &      $\scale{\vb}$ &  $\scale{\taugap}$  &     $\scale{\dec}$ &           $\scale{\bdec}$ &         $\scale{\mm}$ &           $\scale{\bias}$ &          $\scale{\biasO}$ &            $\scale{\lm}$ \\
\midrule
\elecA  &  23 &  0.60 &            \phantom{-}1.2 &    \phantom{-}\textbf{2.3} &            -0.1 &              -0.1 &               \phantom{-}0.6 &                    \phantom{-}0.5 &            \phantom{-}0.1 &            \phantom{-}0.7 &            \phantom{-}0.0 &            \phantom{-}0.2 &              \phantom{-}0.0 &   \phantom{-}\textbf{2.0} \\
\elecB  &  25 &  0.60 &            \phantom{-}0.3 &             \phantom{-}1.5 &            -0.7 &              -0.7 &              -0.1 &                   -0.6 &            \phantom{-}0.1 &            \phantom{-}0.8 &            \phantom{-}0.0 &            \phantom{-}0.2 &              \phantom{-}0.0 &            \phantom{-}1.9 \\
\elecC  &  25 &  0.60 &           -1.3 &             \phantom{-}0.0 &            -1.8 &              -1.8 &              -1.5 &                   -1.8 &            \phantom{-}0.0 &            \phantom{-}0.8 &            \phantom{-}0.0 &            \phantom{-}0.2 &              \phantom{-}0.0 &            \phantom{-}1.8 \\
\elecD  &  10 &  0.64 &  \textbf{-2.9} &            -1.0 &   \textbf{-3.3} &     \textbf{-3.4} &     \textbf{-3.3} &          \textbf{-3.3} &             &             &            \phantom{-}0.0 &            \phantom{-}0.0 &              \phantom{-}0.0 &             \\\midrule
\elecE  &  12 &  0.52 &  \textbf{-3.9} &   \textbf{-3.4} &   \textbf{-2.8} &     \textbf{-2.7} &     \textbf{-3.5} &          \textbf{-2.7} &           -0.1 &            \phantom{-}0.1 &            \phantom{-}0.3 &            \phantom{-}0.9 &    \textbf{-2.0} &            \phantom{-}0.3 \\
\elecF  &  10 &  0.51 &  \textbf{-2.4} &   \textbf{-2.1} &            -1.7 &              -1.6 &     \textbf{-2.1} &                   -1.6 &           -1.0 &           -0.7 &           -0.1 &  \textbf{-2.2} &              \phantom{-}0.0 &           -0.5 \\
\elecG  &  10 &  0.52 &           -0.7 &            -0.4 &            -0.7 &              -0.7 &              -0.7 &                   -0.7 &           -0.8 &           -0.8 &  \textbf{-2.1} &           -1.1 &             -1.6 &           -0.5 \\
\elecH  &  10 &  0.52 &           -0.7 &            -0.4 &            -0.7 &              -0.7 &              -0.8 &                   -0.9 &           -1.0 &           -1.2 &  \textbf{-4.4} &           -1.1 &             -1.6 &           -1.4 \\\midrule

\elecI  &  10 &  0.57 &  -2.1 &            \textbf{-1.1} &   -2.1 &     -2.0 &     -2.2 &                   \textbf{-1.8} &  -2.5 &  -2.0 &            \phantom{-}\textbf{0.0} &            \phantom{-}\textbf{0.0} &              \phantom{-}\textbf{0.0} &           \textbf{-1.5} \\
\elecJ  &  10 &  0.44 &  -2.9 &   -3.3 &            \textbf{-1.4} &              \textbf{-1.4} &     -2.3 &                   \textbf{-1.6} &  -2.0 &  -2.7 &  -2.1 &           \textbf{-1.1} &             \textbf{-1.6} &  -3.4 \\
\elecK &  10 &  0.50 &   \phantom{-}2.7 &    \phantom{-}2.5 &             \phantom{-}\textbf{1.8} &               \phantom{-}\textbf{1.7} &      \phantom{-}2.4 &                    \phantom{-}\textbf{1.8} &   \phantom{-}2.2 &   \phantom{-}2.2 &            \phantom{-}\textbf{1.4} &   \phantom{-}2.2 &     \phantom{-}3.2 &   \phantom{-}2.2 \\
\elecL &  10 &  0.30 & -2.7 &   -4.9 &             \phantom{-}\textbf{0.0} &               \phantom{-}\textbf{0.0} &              \textbf{-1.6} &                   \textbf{-0.3} &  -2.0 &  -3.2 &            \phantom{-}\textbf{1.4} &   \phantom{-}\textbf{2.2} &    -2.4 &  -5.0 \\
\bottomrule
\end{tabular}
\label{tab:hypo}
\end{table}
}

\section{Sensitivity}
\label{sec:sens}

As explored in Section~\ref{sec:semantics}, it only makes sense to
pronounce a district plan as fair or not in the context of a given
electoral environment. As the electoral tide ebbs and flows, so will
the performance of the plan. As such, we fully expect the valuations
of a plan to change from election to election. Nonetheless, moderate
swings should typically lead to moderate valuation changes. Or, if
not, we should at least be able to characterize when a given measure
is particularly volatile so as to interpret the results accordingly.

\ifdef{\SUBMIT}{}{
\begin{figure}
  \centering
  \includegraphics[width=1\linewidth]{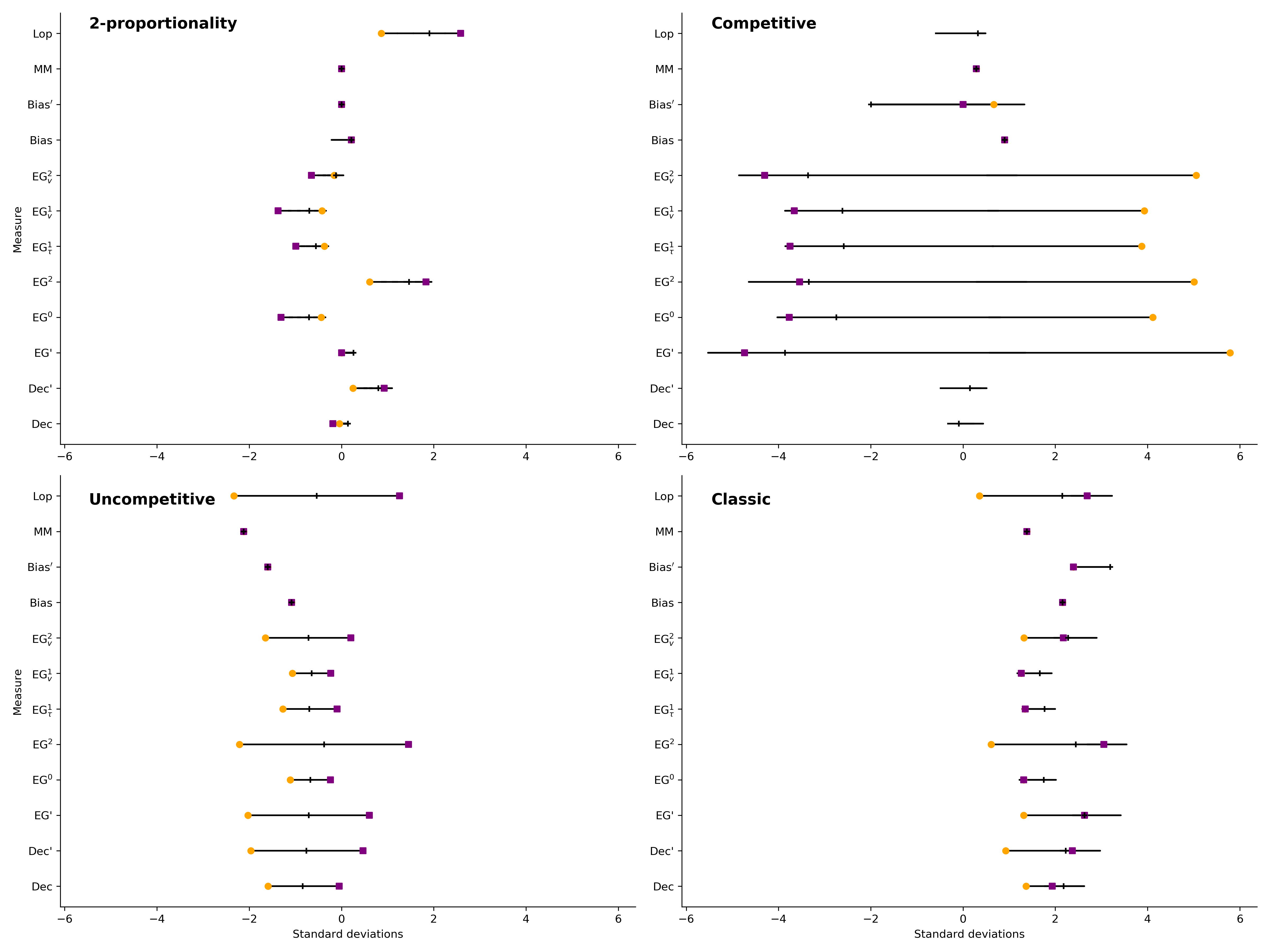}
  \caption{\mycapsens}
  \label{fig:sens}
\end{figure}
}

In Figure~\ref{fig:sens} we illustrate the sensitivity of each measure
to variations in the statewide vote share for four of the hypothetical
elections used in Section~\ref{sec:hypo}. (Plots for the remaining
eight elections are included in the appendix.) For each election, we
apply a series of uniform shifts ranging between -5\% and +5\% and
record the valuation of each measure at each shift. This particular
range has been chosen so as to correspond with the generally accepted
range for competitive elections as being those polling between 45\%
and 55\% for each party. The standardized measure values corresponding
to no shift, -5\% and +5\% are marked by plus signs, orange circles
and purple squares, respectively. The equal vote weights standard is
omitted since it is numerically equivalent to the mean-median
difference. We include $\scale{\bias}$ and $\scale{\mm}$ for reference
even though it is immediate from their definitions that they are
completely insensitive to uniform vote swings. We note that similar
sensitivity analyses have been conducted for various measures using
historical election results --- see, for
example,~\citep{declination,M-S,McDonald-two}. Also, it is worth
mentioning that an alternative approach to sensitivity would be to add
in random noise to the Democratic vote in each district.

Most measures are relatively stable for Election \elecB. The first
exception is the lopsided-means standard, $\scale{\lm}$, which expects
the Democrats' average winning margin to go \emph{down} as their
statewide support increases. Similarly, when the Democratic support
increases, $\scale{\dg}$ views the election as becoming more unfair as
proportionality drifts further from three. The measures $\scale{\va}$
and $\scale{\lossg}$, which are dominated by the losing wasted votes,
have similar issues.

Election \elecE\ highlights the strong sensitivity
the efficiency-gap variants have to the number of seats won. The
measure $\scale{\biasO}$, which is only interested in what happens at
the observed statewide vote share and its complement, runs into
similar trouble that can be traced back to the fact that seat shares
are not distributed evenly across the relevant vote shifts that are
considered.

There is a fair amount of sensitivity for Election \elecF. We see this
as appropriate. As the statewide vote shifts, this election ranges
from one that can be construed as reasonably fair to one in which the
Democrats unquestionably have a strong advantage. The measure
$\scale{\biasO}$ remains completely stable as no districts are flipped
under the vote shifts considered.

The sensitivities for Election \elecK\ are similar to that for Election
\elecG. Again we see moderate sensitivity as appropriate given the
range from an election in which the Republicans have much narrower
winning margins on average to one in which the difference is much more
moderate.

\section{Discussion}
\label{sec:disc}

The investigations of the previous two sections serve to illuminate
various strengths and weaknesses of each measure. We now proceed to
summarize what we have learned. From Table~\ref{tab:hypo}, it is clear
that for the elections chosen, the declination variants perform the
best. These measures are undefined for Election \elecD, but have
neither false positives nor false negatives on the remainder. We also
believe Election \elecL\ is evaluated appropriately by both. The other
measures all have at least three instances of being undefined, having
a false positive, or having a false negative among the twelve
elections. Looking beyond the two versions of the declination, if one
is wishes to minimize false positives, then the mean-median
difference, $\scale{\mm}$, or partisan bias, $\scale{\bias}$, is
probably the best, at least when the statewide vote is close to
even. As these measures look at particular parts of the seats-votes
curve, they may also work better when you have a competitive state in
which the gerrymandering is localized; the other measures, by
averaging over many districts, are less likely to pick up on
gerrymandering in just one or two districts. If one wishes to minimize
false negatives, then $\scale{\eg}$ is a good choice.

The Goldilocks Principle suggests that the measures should have
moderate amounts of sensitivity to both statewide vote share and
competitiveness. Both $\scale{\bias}$ and $\scale{\mm}$ lack any
sensitivity to statewide vote share. On the other hand the variants of
the efficiency gap are all prone to extreme sensitivity to vote share
when there are a large number of competitive races. The declination
and buffered declination both have moderate sensitivity in the
elections considered; this is true of the $\scale{\lm}$ and
$\scale{\biasO}$ as well. So on sensitivity grounds as well we view
these four as the best performers.

The elections considered provide limited data regarding sensitivity to
overall competitiveness. As noted above, the mean-median difference is
perhaps too sensitive to competitiveness as evidenced by the different
evaluations of Elections \elecG\ and \elecH. On the other hand, the
efficiency gap variants perhaps aren't sensitive enough to the highly
competitive nature of Elections \elecE\ and \elecF.

We now discuss the pros and cons of each measure in more detail.

\subsection{Variants of the efficiency gap}
A few of the measures are unusable as measures of partisan
gerrymandering. $\scale{\lossg}$, $\scale{\va}$ and $\scale{\dg}$ are
all measures of proportionality that do not comport with the
historical ``winner's bonus'' that is seen in actuality. It is
unreasonable to adhere to a measure of fairness for which the majority
of historical plans are automatically given to be far from fair. The
final variant of the efficiency gap, $\scale{\vb}$ is perhaps the most
promising, though its evaluation of Election \elecC\ is a mark in its
disfavor. Given its general concurrence with other efficiency gap
variants, especially $\scale{\eg}$ itself, there are not clear
reasons, from the point of view of this article, to consider it or
$\scale{\taugap}$ separately. (We allow, however, that there may be
legal reasons that make $\scale{\vb}$ more attractive.)

\subsection{Mean-median difference and partisan bias}
As noted previously, $\scale{\mm}$ and $\scale{\bias}$ are closely
related. Neither takes into account the fraction of seats won by each
party at the observed statewide vote level (unless each party happens
to have equal statewide support of 50\%). As such, they are completely
insensitive to uniform vote swings.

As such, these measures miss one of, if not \emph{the}, primary goals
of partisan gerrymandering. As shown in the appendix, the outlier
elections for $\scale{\mm}$ are heavily skewed towards ones with few
competitive races. Among hypothetical elections it does reasonably
well in that it marks the proportional and competitive elections as
fair. But it only views the classic gerrymander and inverted
gerrymander as moderately unusual. As discussed in~\citep{declination}
and elsewhere, packing and cracking can occur in such as a way as to
be completely missed by $\scale{\mm}$ when the statewide Democratic
support deviates even moderately from 50\%. It is generally
acknowledged that this measure is not useful when statewide support of
the two parties is unequal.

Partisan bias, in both its variants, performs similarly to
$\scale{\mm}$. $\scale{\bias}$ badly misvalues the partisan advantage
in Election \elecE, but gives what we view as reasonable answers for
the other hypothetical elections. $\scale{\biasO}$ performs similarly,
except that it is Election \elecF\ that it badly misjudges. The large
uniform shift required for elections in which statewide support is far
from even causes issues for $\scale{\bias}$ and, even more so, for
$\scale{\biasO}$.

\subsection{Efficiency gap}
The efficiency gap has as both a strength and a weakness the fact that
it enshrines a historical average as an ideal. By its very definition,
any election that deviates far from this ideal is a good subject for
further investigation. As shown in the appendix (see
Figures~\ref{fig:extremal-comp-DG} and~\ref{fig:extremal-un-dg}), the
elections it tends to flag as outliers \emph{are} unusual in some
respect (e.g., a large number of very competitive seats). Yet there
are also valid reasons for proportionality to deviate from
$2$-proportionality (e.g., a large number of very competitive seats)
and $\scale{\eg}$ is not able to account for these.

\subsection{Declination}
We see the declination and buffered declination as robust measures. A
major advantage of them, in our view, is that they are agnostic about
the slope of the seats-votes curve. They just ask for the differential
responsiveness to be low. As a further point of support,
$\scale{\dec}$ provably increases in absolute value under packing and
cracking, as shown in~\citep{declination}. Disadvantages of
$\scale{\dec}$ and $\scale{\bdec}$ are their inability to provide an
evaluation when one party sweeps the election and their sensitivity
when one party wins only one or two seats, although this latter
property is less true by design for $\scale{\bdec}$.

\subsection{Lopsided means}
The lopsided means measure is predicated on the notion that unequal
average winning vote shares are indicative of unfairness. This is true
when the two parties have exactly equal support statewide: In this
case, having a higher average winning vote share is equivalent to
winning fewer seats. However, the logic behind this test fails even
for moderate swings in statewide support. Average winning margins are
very relevant to the matter of partisan asymmetry, but they must be
interpreted in the context of seats won, as the declination
does. Given the similarities in definition between $\scale{\lm}$ and
$\scale{\dec}$ and the superior results of the latter, we see no
reason to use the lopsided-means measure, especially not in
conjunction with the Student $t$-test (see Section~\ref{sec:lmt}).

\subsection{Lopsided-means test}
\label{sec:lmt}

The lopsided-means test claims to ascertain in a statistically rigorous
manner whether the winning margins of each party could have reasonably
arisen by chance. But the application of this test is highly
problematic. While the venerable $t$-test is a workhorse of
statistics, we do not believe that it is appropriately applied in the
lopsided-means test.

Here, for example, is an instance in which a uniform shift in the
direction of one party causes the average winning vote share of the
new majority party goes up. Begin with evenly matched parties in a
state with 25 districts and with district-level Democratic support
ranging linearly from $0.3$ to $0.7$. The Democrats and Republicans
each have average winning vote shares equal to $0.61$. If there is a
uniform swing of $+5\%$ in the Democrats favor, we end up with
district-level Democratic support ranging linearly from $0.35$ to
$0.75$:
\begin{equation*}
  0.35, 0.36\bar{6}, 0.38\bar{3}, 0.41\bar{6},\ldots, 0.73\bar{3},0.75.
\end{equation*}
The Democrats now win $15$ seats (60\%) with an average winning vote
share of $0.63$ while the Republicans win $10$ seats with an average
winning vote share of $0.58$. The Student $t$-test results in a
$z$-value of $2.16$ and a $p$-value of approximately $0.04$. As this
is less than $0.05$, according to the lopsided-means test, this is
statistically significant and indicative of gerrymandering. But it is
hard to see how this linear vote distribution, shifted only five
points, could suddenly be strongly unfair, especially since the
responsiveness has a historically reasonable value of two. As such, we
view this as a false positive. Similar examples can be constructed in
which a symmetric distribution is shifted slightly in the Democrats'
favor, yet in which their average winning margin decreases.

Nor is it difficult to construct examples that are false
negatives. Consider now a hypothetical district plan with Democratic
support among 15 districts given by
\begin{equation*}
  0.4,0.41,0.42,0.43,0.44,0.45,0.46,0.47,0.48,0.49,0.52,0.55,0.58,0.61,0.64.
\end{equation*}
Democratic support statewide is $0.49$, yet the Democrats win only one
third of the seats in this scenario. The average winning shares in
this case are $0.585$ (Democrats) and $0.55$ (Republicans). The
$t$-test returns a $z$-value of $1.25$ with a $p$-value of $0.23$. Yet
Postulate 2 applies. While this vote distribution is less unfair than
recent House distributions for Pennsylvania or North Carolina, it is
certainly reminiscent of them.

Another issue with the lopsided means test arises when it is applied
to an election with imputed vote values. While any measure will vary
according to the imputation strategy, the $t$-test claims a precision
that does not capture the true uncertainty involved.  As an example,
we consider recent Wisconsin state assembly elections as considered
in~\citep[Fig. 5A]{Wang-elj}. For the 2010 election, for instance,
Wang indicates a difference in average winning vote share of about one
percent that is not statistically significant according to the
$t$-test ($t$-statistic of $-0.56$ and $p$-value of $0.57$). This
value is computed using imputed vote shares of $0.75$ and $0.25$ for
the winner and loser, respectively in each uncontested district. But
since a significant fraction of the races in the election were
uncontested in 2010 --- 31 out of 99 --- a different imputation
strategy can yield markedly different results. For instance, when we
impute vote shares for uncontested races using a hierarchical model as
in~\citep{declination}\footnote{The model consisted of random
  district, state and year effects as well as indicator variables for
  which party won the election and for which party, if any, had an
  incumbent.}, the difference in winning vote shares becomes about 3\%
and the $t$-statistic is computed as $-2.15$ which is statistically
significant with a $p$-value of $0.03$.  For elections with few
uncontested seats, the exact imputation strategy implied will be of
minor importance. But when this is not the case, the imputed support,
depending on how it is done, can lead to more confidence that a
difference is statistically significant than is warranted.

\subsection{Equal vote weights test}
The equal vote weights test is a reasonable proposal assuming the
validity of the mean-median measure. It takes a conservative-sounding
approach of restricting its attention to only elections that are
anti-majoritarian. As its creators acknowledge~\citep{McDonald-two},
this restricts its applicability. Fundamentally, though, it does not
appear to flag particularly interesting elections (see
Figure~\ref{fig:extremal-both-EVW} in Appendix). An anti-majoritarian
outcome is unquestionably a red flag, but it is not clear how the
inclusion of the mean-median difference improves our ability to
identify likely gerrymanders. If one wishes to only impose an
anti-majoritarian requirement, we believe it would be more effectively
done --- on a quantitative level --- with the declination rather than
with the mean-median difference.  We also note that with the endogenous data we
use in this article, only about ten percent of elections with at least
seven districts result in an anti-majoritarian outcome. This measure is
probably best used in elections for which both parties enjoy equal
statewide support and for which any gerrymandering is local in
nature. Another scenario in which it is more likely to be useful is
when there are many seats.

\section{Conclusion}
\label{sec:conc}

This article has focused on measures that attempt to ascertain how
consistent a given vote distribution is with that produced by a
partisan gerrymander. We conclude that the declination,
$\scale{\dec}$, and buffered declination, $\scale{\bdec}$, are the
most successful at avoiding false positives and false negatives among
the elections considered. Their only failing is that they are
undefined on elections in which one party does not win any seats. They
also appear to be relatively robust with regards to sensitivity. And
while $\scale{\bdec}$ may perform better when one party wins only a
seat or two, for simplicity, $\scale{\dec}$ is probably the best
choice overall.

As noted in detail in the discussion, each of the other measures has
serious drawbacks. When there are few competitive elections,
$\scale{\eg}$ and $\scale{\vb}$ might be useful. However, when
statewide support is evenly split, $\scale{\mm}$ and $\scale{\bias}$
might be useful, especially when any alleged gerrymandering is
localized to only a few districts. When one is particularly interested
in anti-majoritarian outcomes, $\scale{\evw}$ might be a good choice.
We do not see any scenario in which $\scale{\dg}$, $\scale{\lossg}$,
$\scale{\va}$, $\scale{\taugap}$ or $\scale{\lm}$ is worth employing.

The framework we employ is a narrow one that intentionally ignores
many of the complexities arising in the partisan gerrymandering
problem. By limiting the scope of what we ask of our measures, we can
do a better job of identifying the strengths and weaknesses of each
measure as well as what election characteristics are likely to lead to
false positives and false negatives. Such errors are inevitable and
must be well understood to be dealt with. Once these quantitative
issues are carefully addressed, it becomes easier to create reliable
ensemble measures (see~\citep{Cain-Cho} for a related discussion in the
context of computer simulations) or to tackle the much larger task of
creating a manageable standard for gerrymandering.

For most of the measures we consider, perturbations to the statewide
vote translate into changes in the evaluations of the measures, as
shown in Section~\ref{sec:sens}. But too much sensitivity can cause
trouble when one is trying to ascertain whether a particular value is
meaningful. In~\citep{Mcgann-america}, the author uses a measure of partisan
symmetry proposed in~\citep{GelKin94} that averages the bias over a
range of Democratic vote shares between 45\% and 55\% (see discussion
in~\citep{Nagle1}). Similar averaging procedures are an obvious way to
increase confidence that a slight change in statewide support wouldn't
lead to a drastically different evaluation by any given measure. Such
techniques could easily be implemented for most of the measures we consider
in this article.

There are many other future directions for study. An obvious direction
is to focus more directly on historical data (as is done briefly in
Table~\ref{tab:outliers} and the appendix). By pronouncing certain
historical district plans as fair or not fair, one could perform an
analysis similar to the one we have performed in this paper. For
example, if the elections in Table~\ref{tab:outliers} were
independently classified as fair or unfair, we would have additional
evidence regarding the utility of the measures considered. Such an
analysis, especially one that took into account unequal turnout among
districts, would bring even more confidence in the applicability of
any conclusions to future elections. A better understanding of how
measures react to different numbers of districts would also be
helpful. Many state legislatures have more than one hundred seats. The
seats-votes curve in such a situation is much closer to a continuous
curve than to the step function it typically is for congressional
elections. Analyses that focus on local district plans or that
consider how these measures might apply to multi-member districts
would also be useful. Beyond that we broaden into matters such as how
partisans are distributed geographically. Such issues are central to
how we should interpret the valuations we get as we move towards
making real-world decisions using these measures. Inherent advantages
from geography, for example, appear to be very real~\citep{chen}, but
if they are modest, then it may be possible in most cases to take
extreme values of a given measure at face value.

While we believe quantitative partisan gerrymandering measures to be
important for solving the partisan gerrymandering problem, they form
only one piece of the puzzle. In particular, we expect computer
simulations (such as found in~\citep{Mattingly17}) and compactness
metrics to continue to play a central role. Combining all of these
techniques into a single robust tool would be of great benefit. Of
course, exactly how such a tool would be used is still an open
question. The US Supreme Court's recent opinion in \emph{Gill
  \emph{v.}\ Whitford}~\citep{gill-new} did little to clarify whether
partisan gerrymanders are unconstitutional. If they are, then it is
reasonable to suppose that partisan-gerrymandering measures will
provide supporting evidence even if they are not the basis of the
constitutional claim itself. And, as we have already mentioned, we
expect these measures to be useful for many different groups during any
redistricting process. The better choice we make for which measures to
use, the more useful they will be.

\section{Data collection and Statistical methods}
\label{sec:mm}

The US state legislature election data up through 2010 comes
from~\citep{carsey}. We only included data on the lower house of the
state legislature when two houses exist. The US congressional data
through 2014 was provided by~\citep{jacobson}. Data for 2016
congressional races were taken from
Wikipedia~\citep{wiki:2016}. See~\citep{redact} for details of data
and imputation strategy.

The election data was analyzed using the python-based
SageMath~\citep{sage}.
\ifdef{\SUBMIT}{}{All non-library code may be found at~\citep{tarball}.}
Python packages employed were pyStan~\citep{pystan} for implementing a
multilevel model to impute votes in uncontested races;
Matplotlib~\citep{matplotlib} and Seaborn~\citep{seaborn} for plotting
and visualization; and SciPy~\citep{scipy} for statistical methods.
\ifdef{\SUBMIT}{}{The data reported in this article are archived at~\citep{redact}.}

\section{Acknowledgments} 
\ifdef{\SUBMIT}{Redacted.}
{
This work was partially supported by a grant from the Simons
Foundation (\#429570). The author is especially indebted to
Gary C. Jacobson for sharing his data on US Congressional
elections. The author thanks Jeff Buzas for helpful conversations.
}

\bibliography{gerrymandering}

\begin{thebibliography}{46}
\expandafter\ifx\csname natexlab\endcsname\relax\def\natexlab#1{#1}\fi
\expandafter\ifx\csname url\endcsname\relax
  \def\url#1{{\tt #1}}\fi
\expandafter\ifx\csname urlprefix\endcsname\relax\def\urlprefix{URL }\fi

\bibitem[{Baas \& McAuliffe(2017)}]{mcauliffe}
Baas, K., \& McAuliffe, C. (2017).
\newblock
  \url{https://github.com/ColinMcAuliffe/UnburyTheLead/raw/master/ELJ/EmpiricalBayes/EmpiricalBayes.pdf}.

\bibitem[{Bangia et~al.(2017)Bangia, Graves, Herschlag, Kang, Luo, Mattingly,
  \& Ravier}]{Mattingly17}
Bangia, S., Graves, C.~V., Herschlag, G., Kang, H.~S., Luo, J., Mattingly,
  J.~C., \& Ravier, R. (2017).
\newblock Redistricting: Drawing the line.
\newblock {\em ArXiv e-prints\/}.
\newblock \url{http://arxiv.org/abs/1704.03360}.

\bibitem[{Bernstein \& Duchin(2017)}]{MoonMira}
Bernstein, M., \& Duchin, M. (2017).
\newblock A formula goes to court: Partisan gerrymandering and the efficiency
  gap.
\newblock {\em Notices of the AMS\/}, {\em 64\/}(9), 1020--1024.

\bibitem[{Best et~al.(2017)Best, Donahue, Krasno, Magleby, \&
  McDonald}]{McDonald-two}
Best, R.~E., Donahue, S.~J., Krasno, J., Magleby, D.~B., \& McDonald, M.~D.
  (2017).
\newblock Considering the prospects for establishing a packing gerrymandering
  standard.
\newblock {\em Elect. Law J.\/}, {\em 17\/}(1).

\bibitem[{Best et~al.(Mar 2018)Best, Donahue, Krasno, Magleby, \&
  McDonald}]{BestResponse}
Best, R.~E., Donahue, S.~J., Krasno, J., Magleby, D.~B., \& McDonald, M.~D.
  (Mar 2018).
\newblock Authors' response --- values and validations: Proper criteria for
  comparing standards for packing gerrymanders.
\newblock {\em Election Law Journal\/}, {\em 17\/}(1), 82--84.

\bibitem[{Cain et~al.(2018)Cain, Cho, Liu, \& Zhang}]{Cain-Cho}
Cain, B.~E., Cho, W. K.~T., Liu, Y.~Y., \& Zhang, E.~R. (2018).
\newblock A reasonable bias approach to gerrymandering: using automated plan
  generation to evaluate redistricting proposals.
\newblock {\em William \& Mary Law Review\/}, {\em 59\/}(5).

\bibitem[{Chatterjee et~al.(2018)Chatterjee, DasGupta, Palmieri, Al{-}Qurashi,
  \& Sidiropoulos}]{chatterjee}
Chatterjee, T., DasGupta, B., Palmieri, L., Al{-}Qurashi, Z., \& Sidiropoulos,
  A. (2018).
\newblock Alleviating partisan gerrymandering: can math and computers help to
  eliminate wasted votes?
\newblock {\em CoRR\/}, {\em abs/1804.10577\/}.
\newline\urlprefix\url{http://arxiv.org/abs/1804.10577}

\bibitem[{Chen \& Rodden(2013)}]{chen}
Chen, J., \& Rodden, J. (2013).
\newblock Unintentional gerrymandering: Political geography and electoral bias
  in legislatures.
\newblock {\em Quart. J. of Pol. Sci.\/}, {\em 8\/}, 239--269.

\bibitem[{Cho(2017)}]{cho-upenn}
Cho, W. K.~T. (2017).
\newblock Measuring partisan fairness: How well does the efficiency gap guard
  against sophisticated as well as simple-minded modes of partisan
  discrimination?
\newblock {\em University of Pennsylvania Law Review\/}, {\em 166\/}(1),
  1263--1321.

\bibitem[{Cover(April 2018)}]{Cover}
Cover, B.~P. (April 2018).
\newblock Quantifying partisan gerrymandering: An evaluation of the efficiency
  gap proposal.
\newblock {\em Stanford Law Review\/}, {\em 70\/}(4), 1131--1233.

\bibitem[{Gannon(June 20, 2016)}]{lewis}
Gannon, P. (June 20, 2016).
\newblock Political gerrymandering nears `too far'.
\newblock {\em Elizabeth City Daily Advance\/}.
\newblock Accessed online, July 12, 2017. Syndicated column.
\newline\urlprefix\url{http://www.dailyadvance.com/Other-Views/2016/06/20/Political-gerrymandering-nears-too-far.html}

\bibitem[{Gelman \& King(1994{\natexlab{a}})}]{GK}
Gelman, A., \& King, G. (1994{\natexlab{a}}).
\newblock Enhancing democracy through legislative redistricting.
\newblock {\em Am. Political Sci. Rev.\/}, {\em 88\/}, 541--559.

\bibitem[{Gelman \& King(1994{\natexlab{b}})}]{GelKin94}
Gelman, A., \& King, G. (1994{\natexlab{b}}).
\newblock A unified method of evaluating electoral systems and redistricting
  plans.
\newblock {\em American Journal of Political Science\/}, {\em 38\/},
  514{\textendash}554.

\bibitem[{Gill()}]{gill-new}
Gill (2018).
\newblock \emph{Gill \emph{v.}\ Whitford,} 585 U.S. \underline{\ \ \ } (2018).

\bibitem[{Hunter(2007)}]{matplotlib}
Hunter, J.~D. (2007).
\newblock Matplotlib: A 2d graphics environment.
\newblock {\em Computing In Science \& Engineering\/}, {\em 9\/}(3), 90--95.

\bibitem[{Jacobson(2017)}]{jacobson}
Jacobson, G.~C. (2017).
\newblock Private communication.

\bibitem[{Jones et~al.(2001--)Jones, Oliphant, Peterson et~al.}]{scipy}
Jones, E., Oliphant, T., Peterson, P., et~al. (2001--).
\newblock {SciPy}: Open source scientific tools for {Python}.
\newblock \url{http://www.scipy.org/} [accessed 2017-02-08].

\bibitem[{Klarner et~al.(2013-01-11)Klarner, Berry, Carsey, Jewell, Niemi,
  Powell, \& Snyder}]{carsey}
Klarner, C., Berry, W.~D., Carsey, T., Jewell, M., Niemi, R., Powell, L., \&
  Snyder, J. (2013-01-11).
\newblock State legislative election returns (1967-2010).
\newblock ICPSR34297-v1.
\newline\urlprefix\url{http://doi.org/10.3886/ICPSR34297.v1}

\bibitem[{Levitt(2018)}]{loyola:nc}
Levitt, J. (2018).
\newblock North carolina.
\newblock Accessed 12-December-2017.
\newline\urlprefix\url{http://redistricting.lls.edu/states-NC.php}

\bibitem[{McGann et~al.(2016)McGann, Smith, Latner, \& Keena}]{Mcgann-america}
McGann, A.~J., Smith, C.~A., Latner, M., \& Keena, A. (2016).
\newblock {\em Gerrymandering in America: The House of Representatives, the
  Supreme Court, and the Future of Popular Sovereignty\/}.
\newblock Cambridge University Press.

\bibitem[{McGann et~al.(December 2015)McGann, Smith, Latner, \&
  Keena}]{McGann-elj}
McGann, A.~J., Smith, C.~A., Latner, M., \& Keena, A.~J. (December 2015).
\newblock A discernable and manageable standard for partisan gerrymandering.
\newblock {\em Election Law Journal\/}, {\em 14\/}(4), 295--311.

\bibitem[{McGhee(2014)}]{McGhee}
McGhee, E. (2014).
\newblock Measuring partisan bias in single-member district electoral systems.
\newblock {\em Legis. Stud. Q.\/}, {\em 39\/}, 55--85.

\bibitem[{McGhee(Augest 11, 2017)}]{McGheeTest}
McGhee, E. (Augest 11, 2017).
\newblock Symposium: The efficiency gap is a measure, not a test.
\newblock {\em SCOTUSblog\/}.
\newblock Accessed online, May 24, 2018.
\newline\urlprefix\url{http://www.scotusblog.com/2017/08/symposium-efficiency-gap-measure-not-test/}

\bibitem[{McGhee(Mar 2018)}]{McGheeRejoinder}
McGhee, E. (Mar 2018).
\newblock Rejoinder to ``{C}onsidering the prospects for establishing a packing
  gerrymandering standard''.
\newblock {\em Election Law Journal\/}, {\em 17\/}(1), 73--82.

\bibitem[{McGhee \& Stephanopoulos(2015)}]{M-S}
McGhee, E., \& Stephanopoulos, N. (2015).
\newblock Partisan gerrymandering and the efficiency gap.
\newblock {\em 82 University of Chicago Law Review\/}, {\em 831\/}.
\newblock 70 pages. U of Chicago, Public Law working Paper No. 493. Available
  at SSRN: https://ssrn.com/abstract=2457468.

\bibitem[{Nagle(2017)}]{Nagle2}
Nagle, J.~F. (2017).
\newblock How competitive should a fair single member districting plan be?
\newblock {\em Elect. Law J.\/}, {\em 16\/}(1), 196--209.

\bibitem[{Nagle(December 2015)}]{Nagle1}
Nagle, J.~F. (December 2015).
\newblock Measures of partisan bias for legislating fair elections.
\newblock {\em Election Law Journal\/}, {\em 14\/}(4), 346--360.

\bibitem[{Redacted()}]{redact}
Redacted (????).
\newblock {R}edacted.

\bibitem[{Stein et~al.(2016)}]{sage}
Stein, W., et~al. (2016).
\newblock {\em {S}age {M}athematics {S}oftware ({V}ersion 7.1)\/}.
\newblock The Sage Development Team.
\newblock {\tt http://www.sagemath.org}.

\bibitem[{Stephanopoulos \& McGhee(2017)}]{MSII}
Stephanopoulos, N.~O., \& McGhee, E.~M. (2017).
\newblock The measure of a metric: The debate over quantifying partisan
  gerrymandering.
\newline\urlprefix\url{https://ssrn.com/abstract=3077766}

\bibitem[{Tapp(2018)}]{Tapp}
Tapp, K. (2018).
\newblock Measuring political gerrymandering.
\newblock {\em ArXiv e-prints\/}.
\newblock \url{http://arxiv.org/abs/1801.02541}.

\bibitem[{Team(2016)}]{pystan}
Team, S.~D. (2016).
\newblock {\em pyStan: the Python interface to Stan, Version 2.14.0.0\/}.
\newblock {\tt http://mc-stan.org}.

\bibitem[{Tufte(1973)}]{tufte}
Tufte, E.~R. (1973).
\newblock The relationship between seats and votes in two-party systems.
\newblock {\em American Political Science Review\/}, {\em 67\/}(3), 540--54.

\bibitem[{Veomett(2018)}]{Veomett}
Veomett, E. (2018).
\newblock The efficiency gap, voter turnout, and the efficiency principle.
\newblock {\em ArXiv e-prints\/}.
\newblock \url{http://arxiv.org/abs/1801.05301}.

\bibitem[{Wang(2016)}]{Wang-elj}
Wang, S. S.-H. (2016).
\newblock Three practical tests for gerrymandering: Application to maryland and
  wisconsin.
\newblock {\em Election Law Journal\/}, {\em 15\/}(4).

\bibitem[{Wang(June 2016)}]{Wang}
Wang, S. S.-H. (June 2016).
\newblock Three tests for practical evaluation of partisan gerrymandering.
\newblock {\em Stanford Law Review\/}, {\em 68\/}, 1263--1321.

\bibitem[{Warrington(2018{\natexlab{a}})}]{tarball}
Warrington, G.~S. (2018{\natexlab{a}}).
\newblock Election data and computer code.
\newblock \url{http://www.cems.uvm.edu/~gswarrin/gerrymandering/}.
\newblock [Online; accessed 23-January-2018].

\bibitem[{Warrington(2018{\natexlab{b}})}]{DecIntro}
Warrington, G.~S. (2018{\natexlab{b}}).
\newblock Introduction to the declination function for gerrymanders.
\newblock {\em ArXiv e-prints\/}.
\newblock \url{http://arxiv.org/abs/1803.04799}.

\bibitem[{Warrington(Mar 2018)}]{declination}
Warrington, G.~S. (Mar 2018).
\newblock Quantifying gerrymandering using the vote distribution.
\newblock {\em Election Law Journal\/}, {\em 17\/}(1).

\bibitem[{Warshaw \& Stephanopoulos(2019)}]{Warshaw}
Warshaw, C., \& Stephanopoulos, N. (2019).
\newblock The impact of partisan gerrymandering on political parties.
\newblock Available at SSRN: \url{https://ssrn.com/abstract=3330695}.

\bibitem[{Waskom et~al.(2014)Waskom, Botvinnik, Hobson, Cole, Halchenko, Hoyer,
  Miles, Augspurger, Yarkoni, Megies, Coelho, Wehner, cynddl, Ziegler,
  diego0020, Zaytsev, Hoppe, Seabold, Cloud, Koskinen, Meyer, Qalieh, \&
  Allan}]{seaborn}
Waskom, M., Botvinnik, O., Hobson, P., Cole, J.~B., Halchenko, Y., Hoyer, S.,
  Miles, A., Augspurger, T., Yarkoni, T., Megies, T., Coelho, L.~P., Wehner,
  D., cynddl, Ziegler, E., diego0020, Zaytsev, Y.~V., Hoppe, T., Seabold, S.,
  Cloud, P., Koskinen, M., Meyer, K., Qalieh, A., \& Allan, D. (2014).
\newblock {S}eaborn: v0.5.0 ({N}ovember 2014).
\newline\urlprefix\url{https://doi.org/10.5281/zenodo.12710}

\bibitem[{Whitford, 2016a()}]{Wisconsin}
Whitford, 2016a (2016).
\newblock \emph{Whitford v. Gill}, No. 15-cv-421, F. Supp. 3d (2016).

\bibitem[{Whitford, 2016b()}]{Griesbach}
Whitford, 2016b (2016).
\newblock \emph{Whitford v. Gill}, No. 15-cv-421, F. Supp. 3d (2016).
  Griesbach, dissenting, 128.

\bibitem[{Wikipedia(2017)}]{wiki:2016}
Wikipedia (2017).
\newblock United {S}tates {H}ouse of {R}epresentatives elections, 2016 ---
  {W}ikipedia{,} the free encyclopedia.
\newblock [Online; accessed 15-February-2017].
\newline\urlprefix\url{\url{https://en.wikipedia.org/wiki/United_States_House_of_Representatives_elections,_2016}}

\bibitem[{Wikipedia(2018{\natexlab{a}})}]{wiki:nc10}
Wikipedia (2018{\natexlab{a}}).
\newblock {U}nited {S}tates {H}ouse of {R}epresentatives elections in {N}orth
  {C}arolina, 2010 --- {W}ikipedia{,} the free encyclopedia.
\newblock Online; accessed Math 30, 2018.
\newline\urlprefix\url{\url{https://en.wikipedia.org/wiki/United_States_House_of_Representatives_elections_in_North_Carolina,_2010}}

\bibitem[{Wikipedia(2018{\natexlab{b}})}]{wiki:nc12}
Wikipedia (2018{\natexlab{b}}).
\newblock {U}nited {S}tates {H}ouse of {R}epresentatives elections in {N}orth
  {C}arolina, 2012 --- {W}ikipedia{,} the free encyclopedia.
\newblock Online; accessed Math 30, 2018.
\newline\urlprefix\url{\url{https://en.wikipedia.org/wiki/United_States_House_of_Representatives_elections_in_North_Carolina,_2012}}

\end{thebibliography}

\bibliographystyle{apa-good}

\clearpage

\newcommand{\beginsupplement}{%
        \setcounter{section}{0}
        \renewcommand{\thesection}{A\arabic{section}}%
        \setcounter{table}{0}
        \renewcommand{\thetable}{A\arabic{table}}%
        \setcounter{figure}{0}
        \renewcommand{\thefigure}{A\arabic{figure}}%
     }

\beginsupplement

\section{Appendix}

\baselineskip 12pt

In this section we present the sensitivity evaluations for the other
eight hypothetical elections. We also collect plots of two different
collections of elections. The first set consists of the most extreme
outliers for each measure, both for all elections in our historical
data set as well as for the subset consisting of those for which the
statewide Democratic support is between 45\% and 55\%. The second set
collects those elections for which the measures disagree most
strongly. In this second set we only consider the measures
$\scale{\bdec}$, $\scale{\mm}$, $\scale{\bias}$ and $\scale{\eg}$.

\clearpage

\subsection{Sensitivity}

\begin{figure}[h]
  \centering
  \includegraphics[width=1\linewidth]{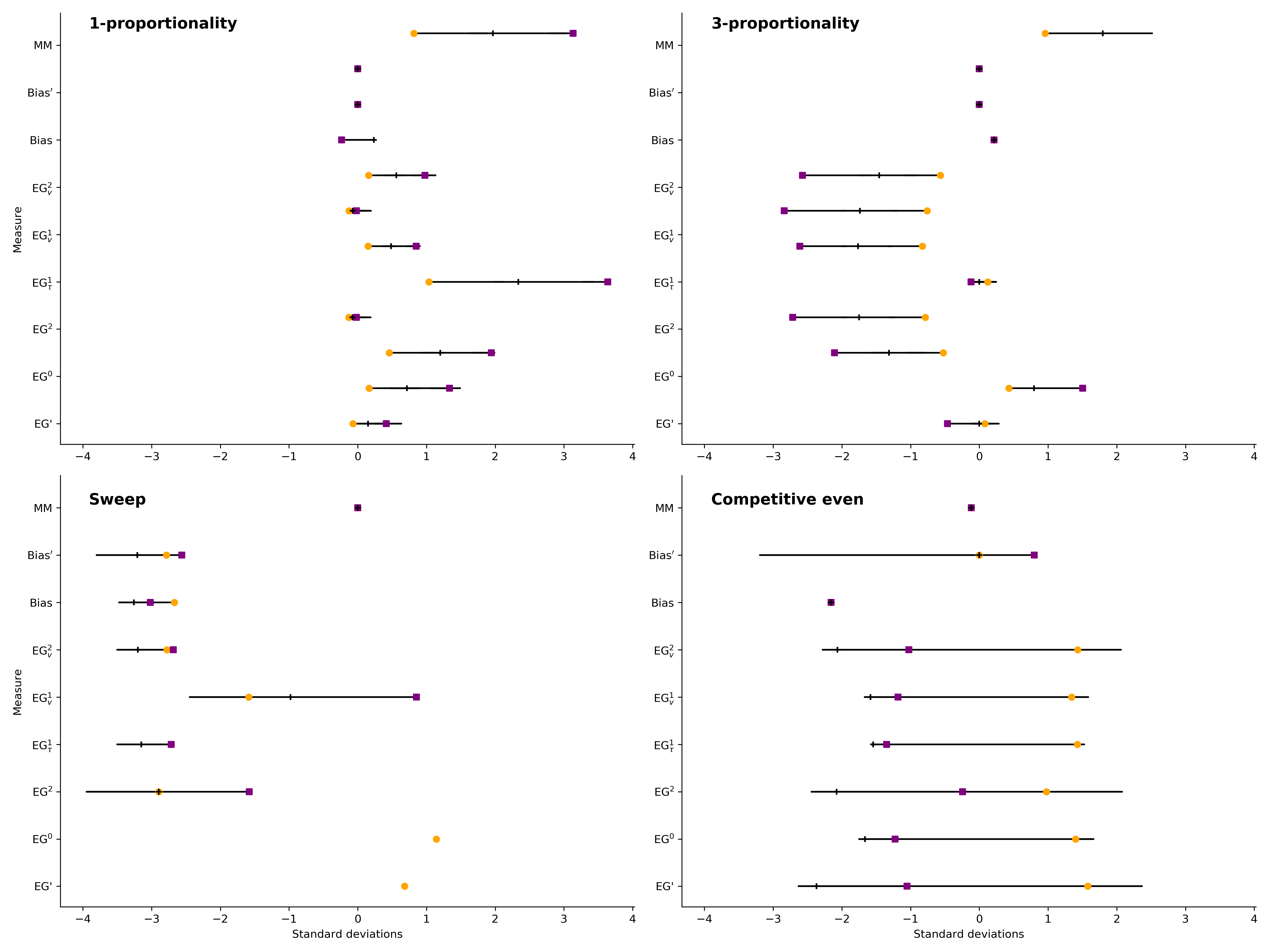}
  \caption{\mycapsens}
  \label{fig:sens2}
\end{figure}

\begin{figure}[h]
  \centering
  \includegraphics[width=1\linewidth]{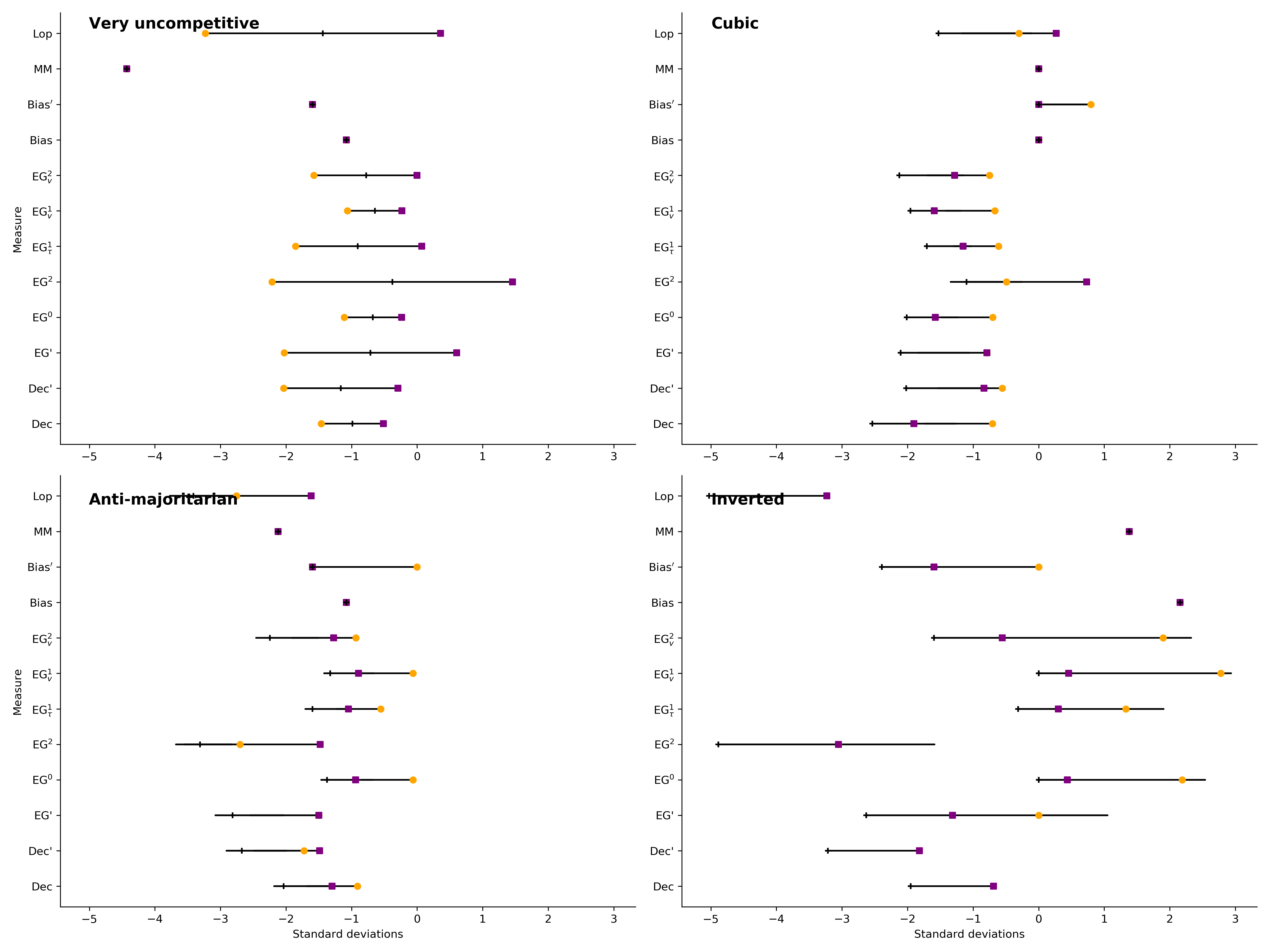}
  \caption{\mycapsens}
  \label{fig:sens3}
\end{figure}

\clearpage

\subsection{Outliers}

\begin{figure}[b]
  \centering
  \includegraphics[width=.9\linewidth]{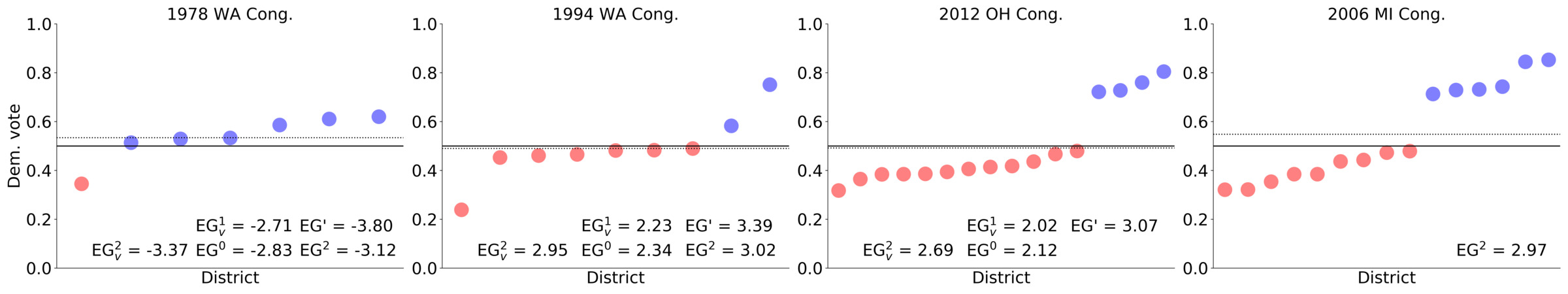}
  \caption{\mycape}
  \label{fig:extremal-comp-DG}
\end{figure}

\begin{figure}[b]
  \centering
  \includegraphics[width=.8\linewidth]{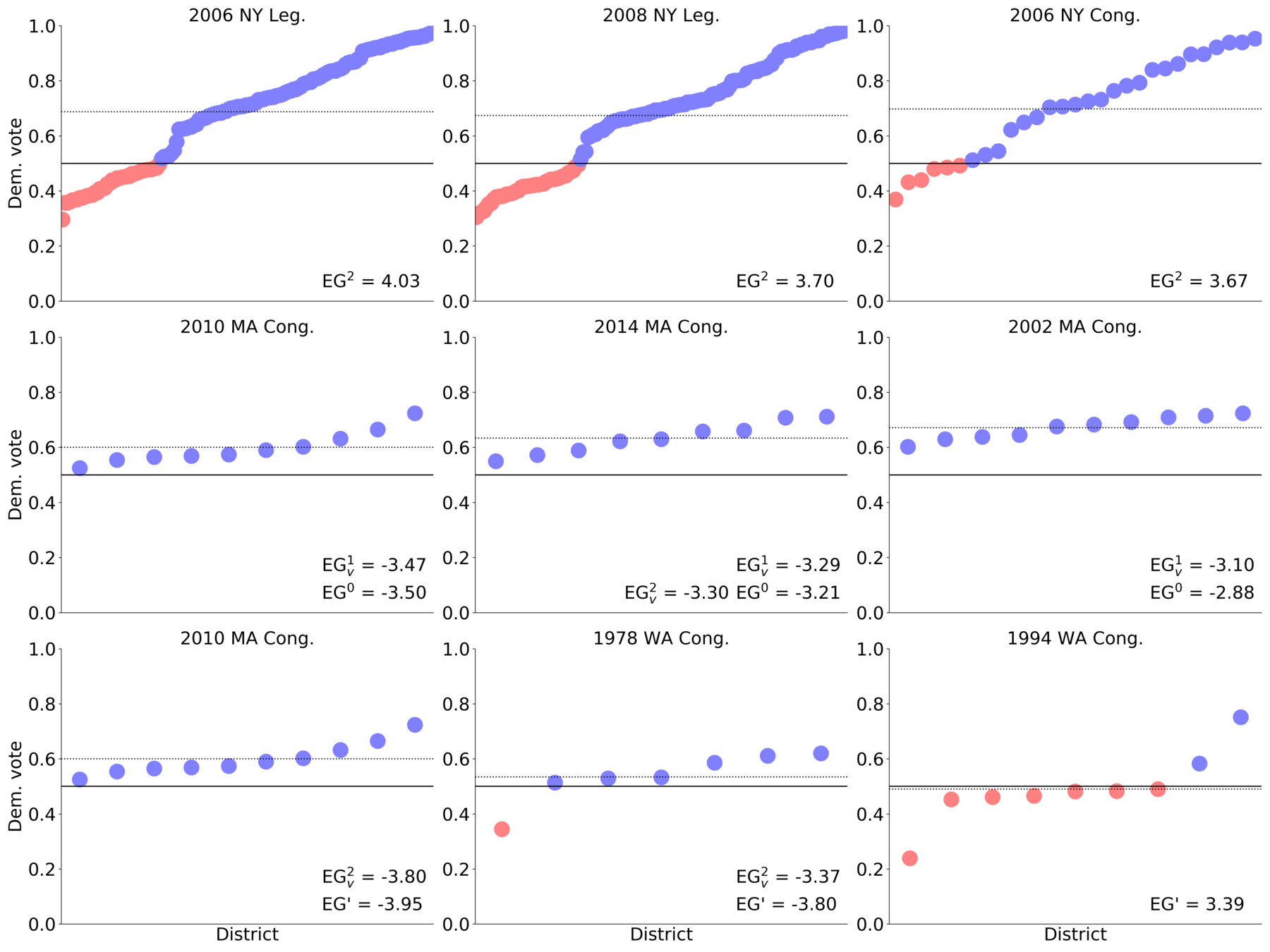}
  \caption{\mycapf}
  \label{fig:extremal-un-dg}
\end{figure}

\begin{figure}
  \centering
  \includegraphics[width=.8\linewidth]{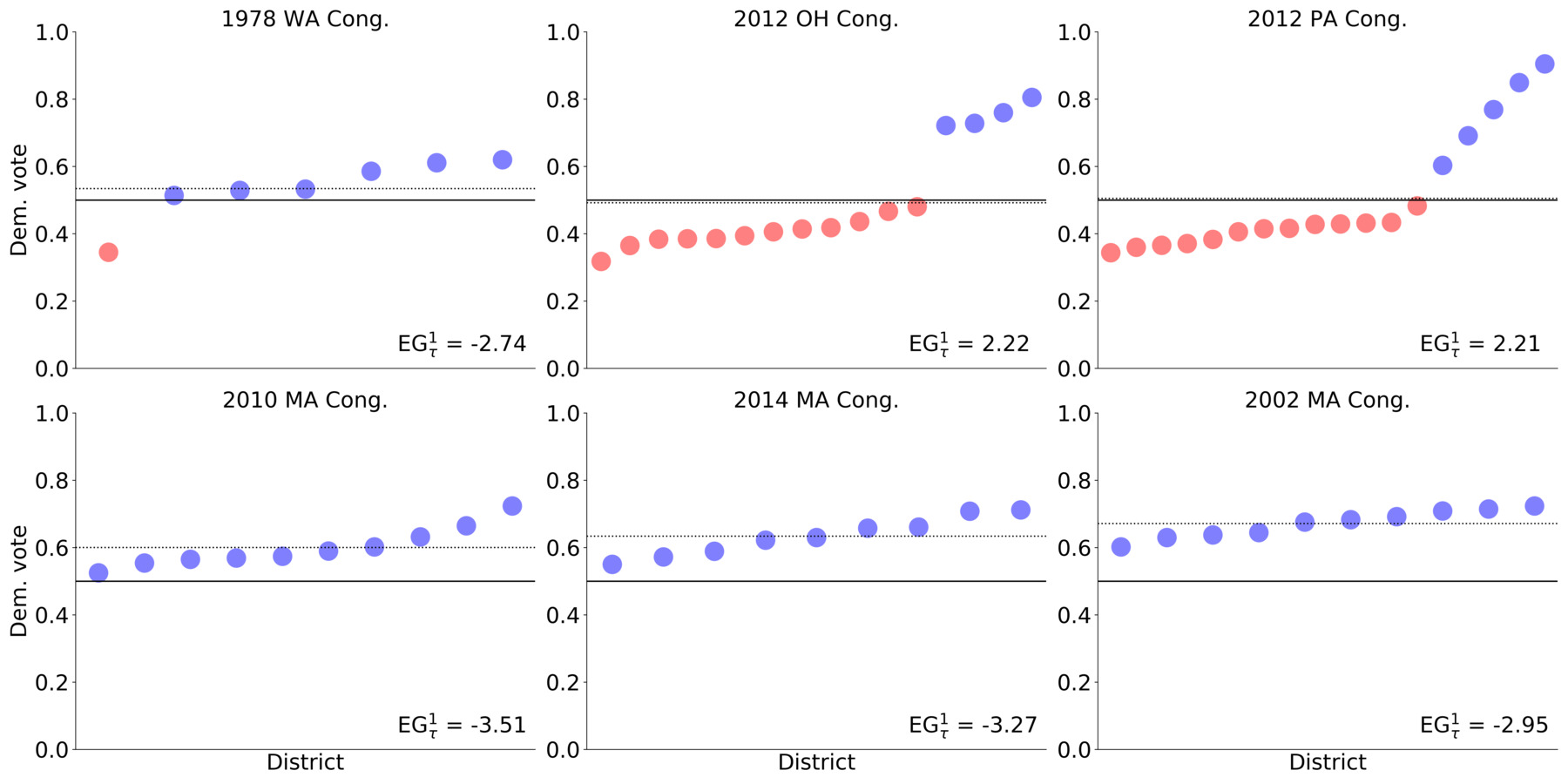}
  \caption{\mycapff}
  \label{fig:extremal-both-Tau}
\end{figure}

\begin{figure}
  \centering
  \includegraphics[width=.8\linewidth]{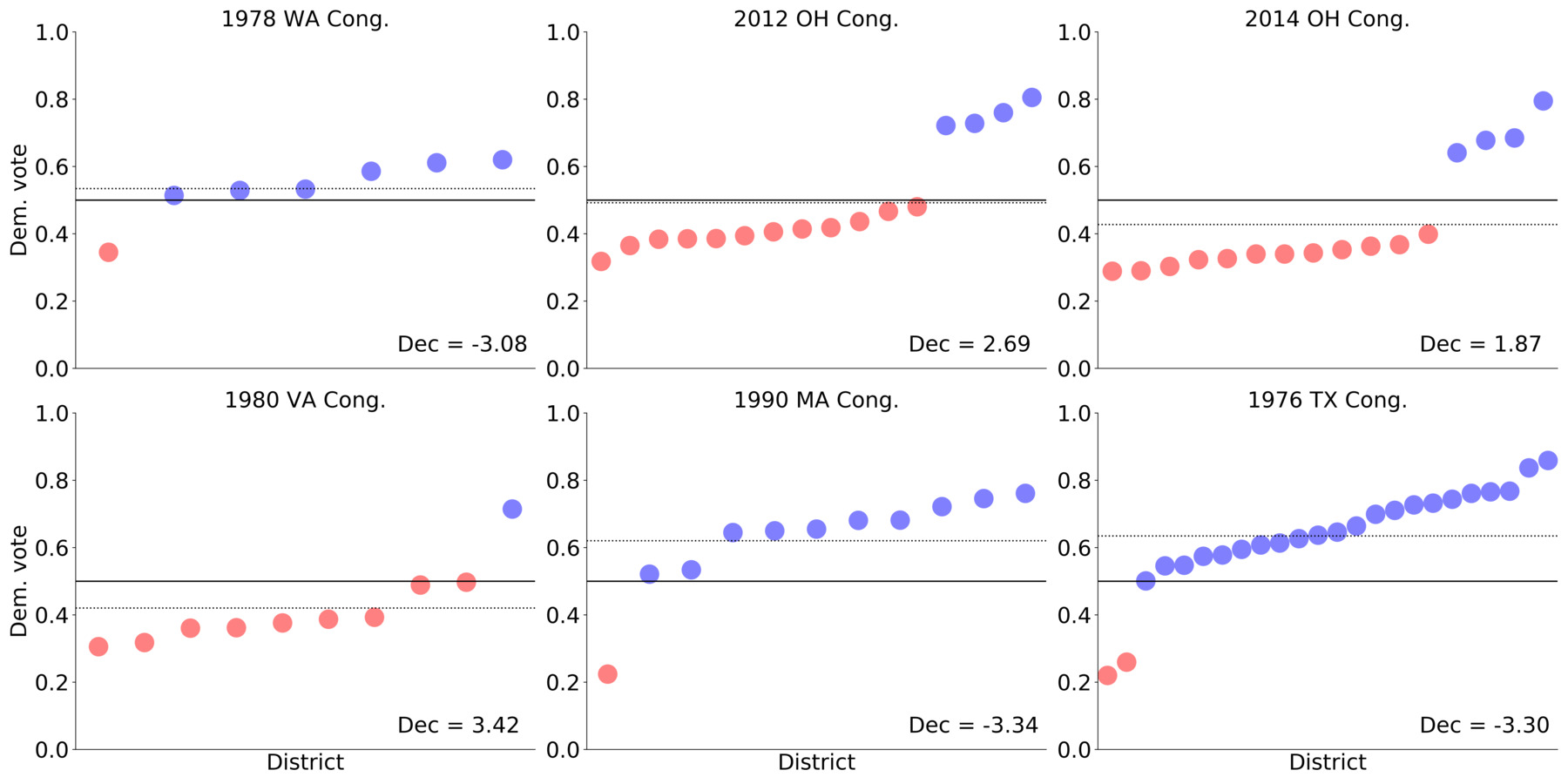}
  \caption{\mycapg}
  \label{fig:extremal-both-Dec}
\end{figure}

\begin{figure}
  \centering
  \includegraphics[width=.8\linewidth]{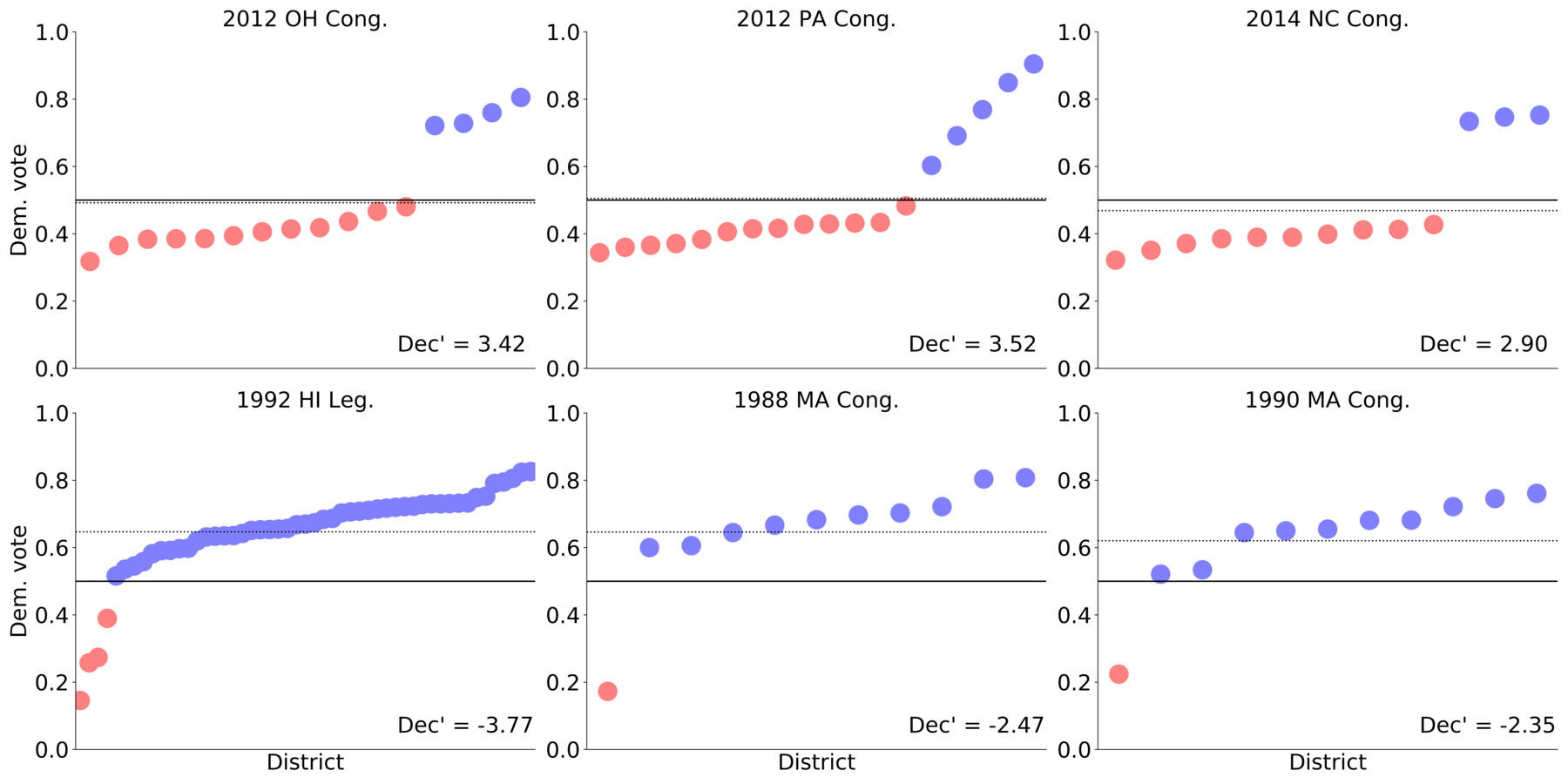}
  \caption{\mycaph}
  \label{fig:extremal-both-BDec}
\end{figure}

\begin{figure}
  \centering
  \includegraphics[width=.8\linewidth]{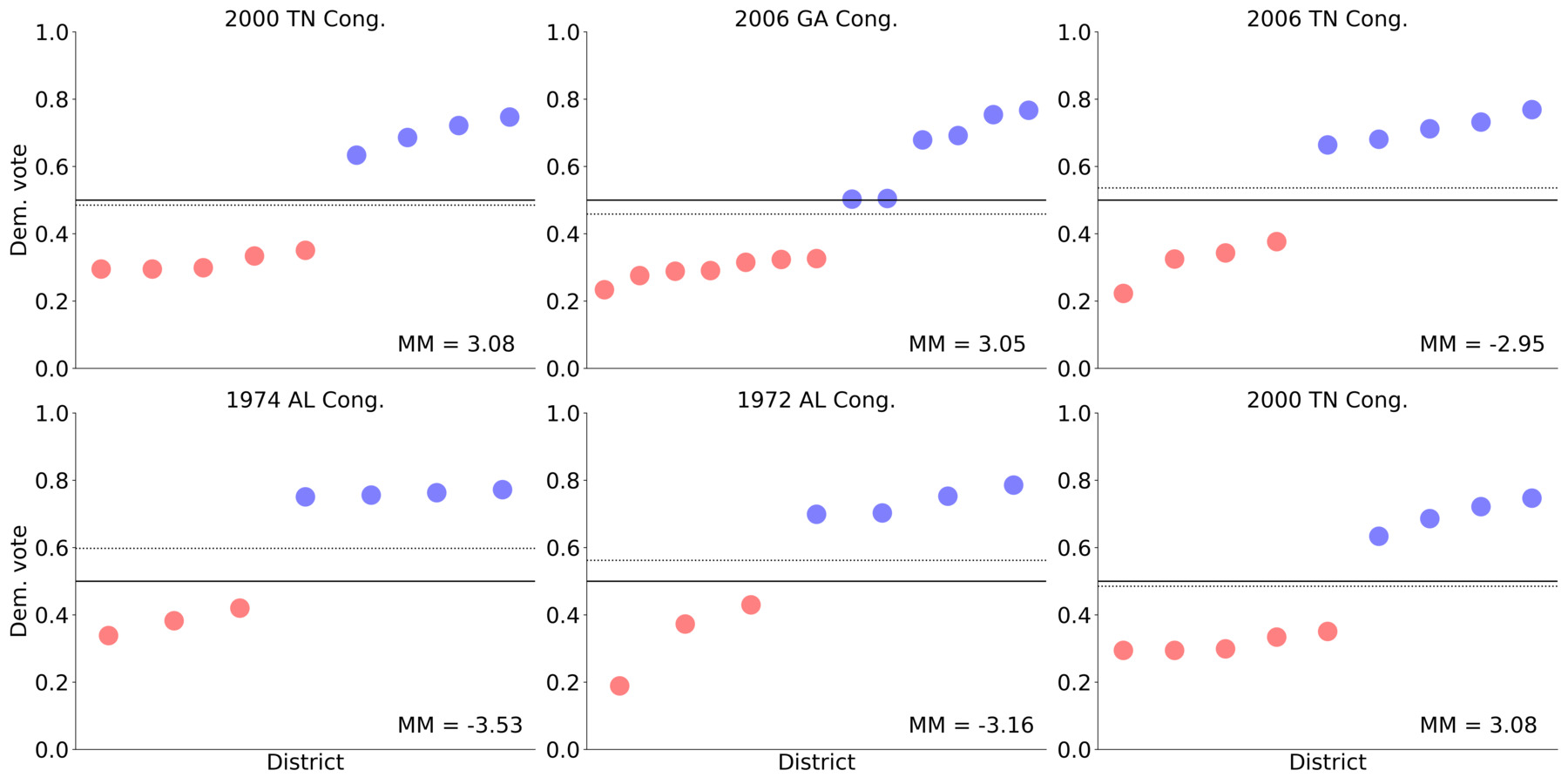}
  \caption{\mycapi}
  \label{fig:extremal-both-MM}
\end{figure}

\begin{figure}
  \centering
  \includegraphics[width=.8\linewidth]{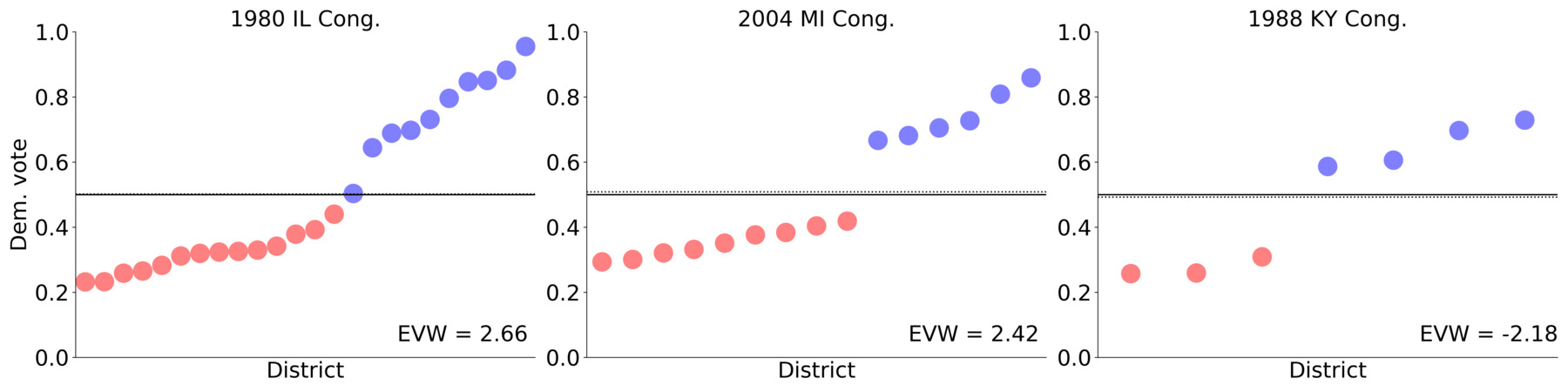}
  \caption{\mycapj}
  \label{fig:extremal-both-EVW}
\end{figure}

\begin{figure}
  \centering
  \includegraphics[width=.8\linewidth]{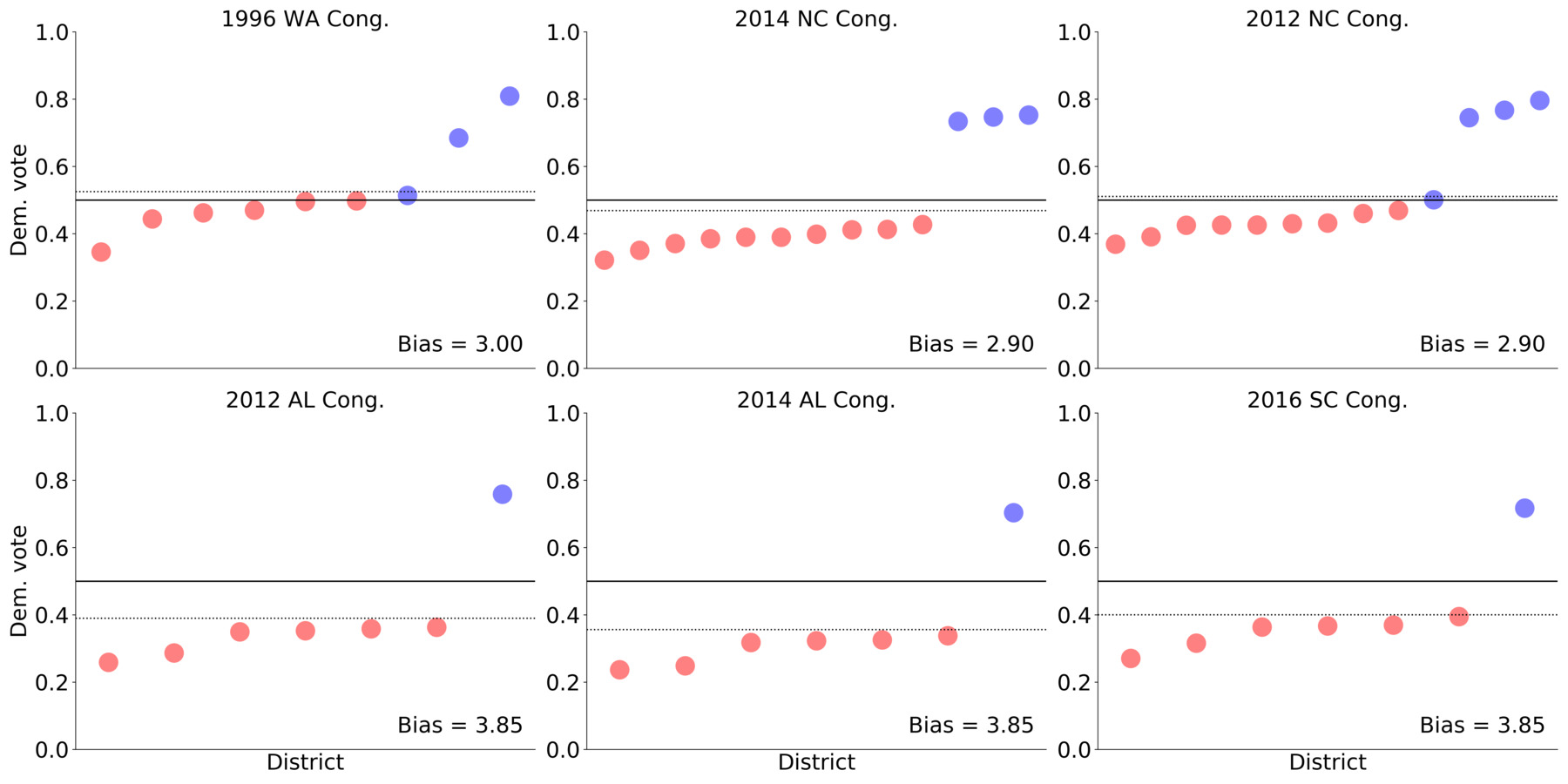}
  \caption{\mycapk}
  \label{fig:extremal-both-Bias}
\end{figure}

\begin{figure}
  \centering
  \includegraphics[width=.8\linewidth]{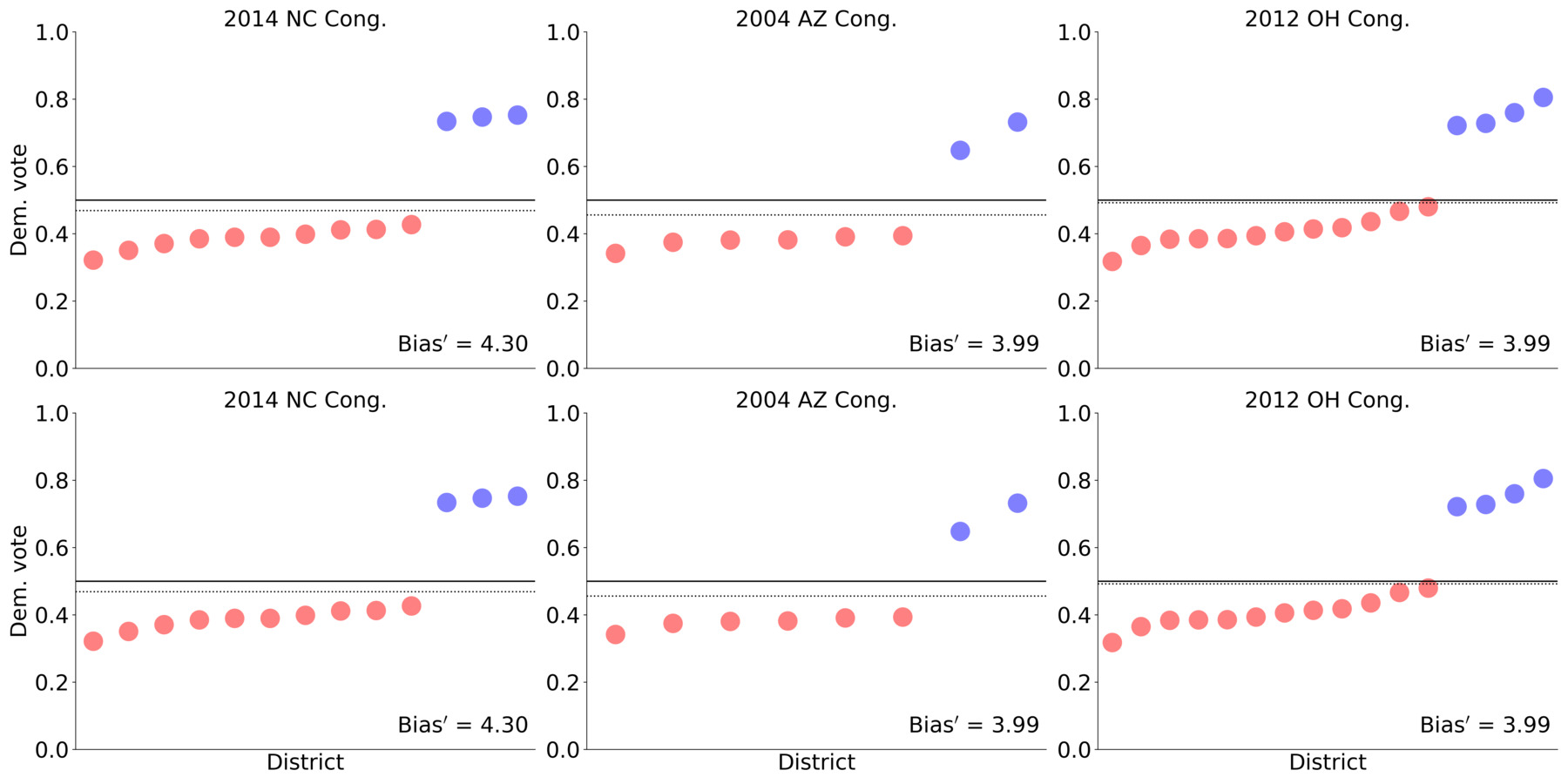}
  \caption{\mycapkii}
  \label{fig:extremal-both-BiasO}
\end{figure}

\begin{figure}
  \centering
  \includegraphics[width=.8\linewidth]{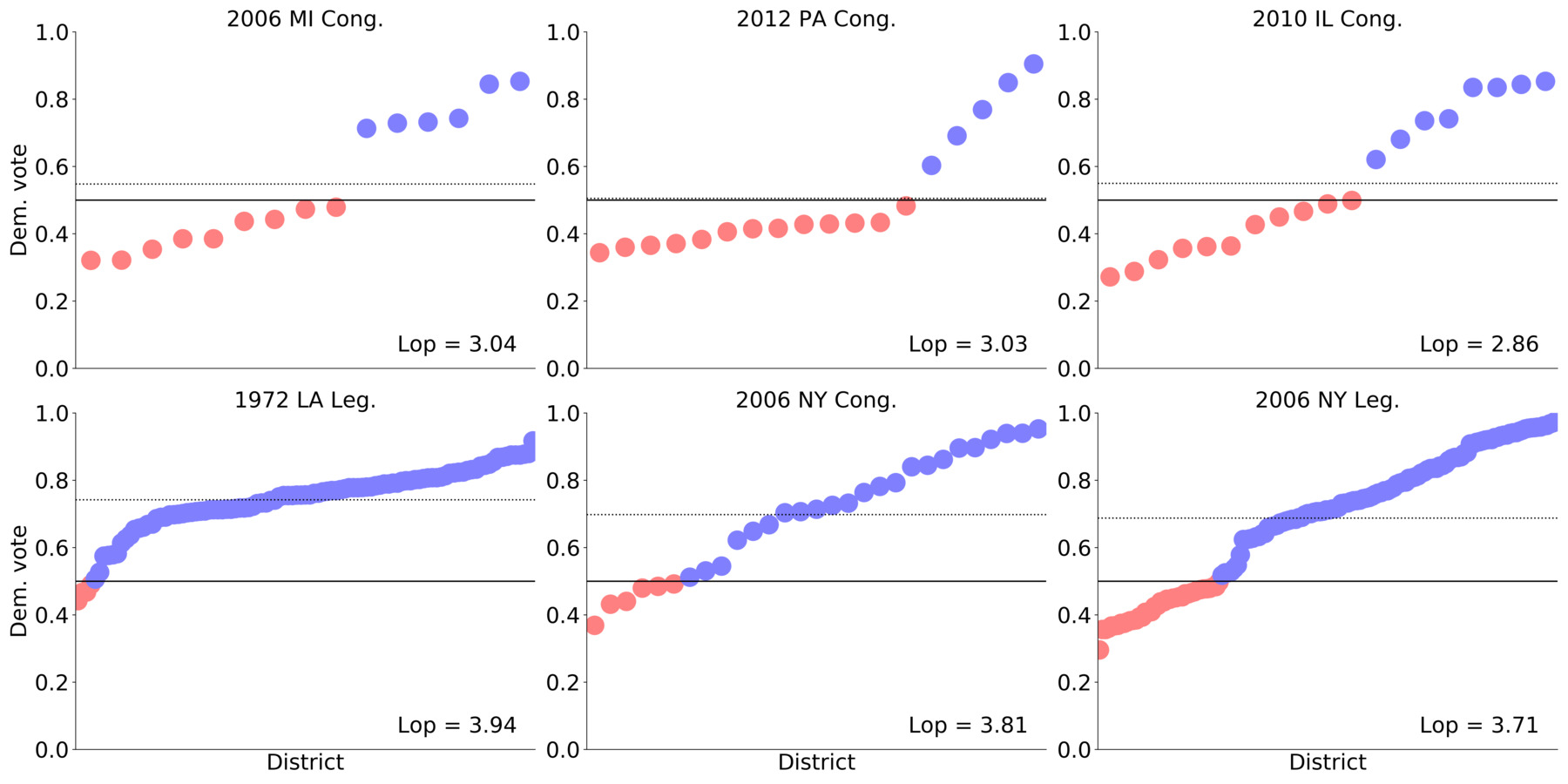}
  \caption{\mycapl}
  \label{fig:extremal-both-Lop}
\end{figure}

\clearpage
\subsection{Greatest pairwise disagreements}
\label{sec:pairs}


\begin{figure}
  \centering
  \includegraphics[width=.8\linewidth]{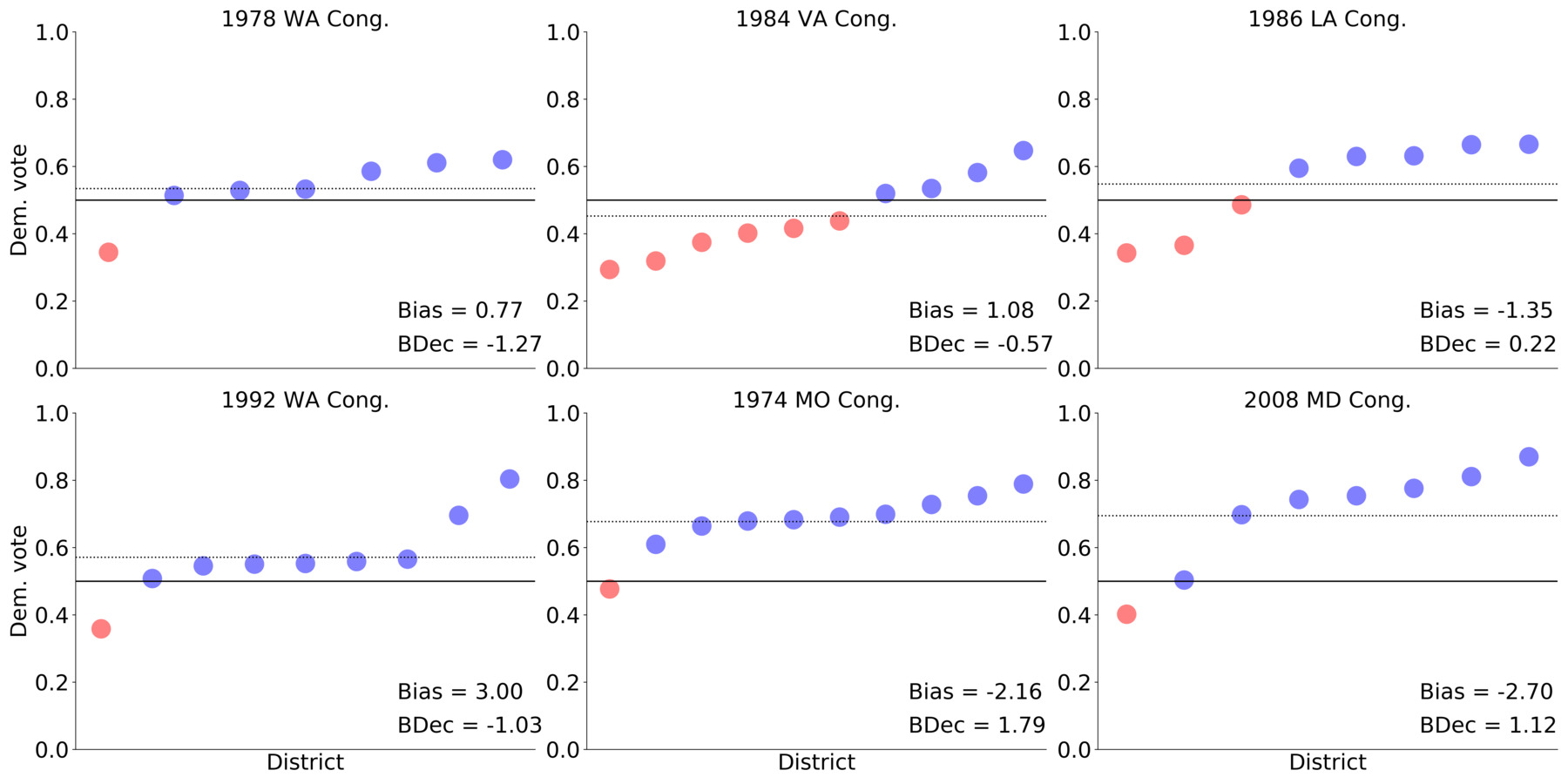}
  \caption{\mycapn}
  \label{fig:worst-BDec-Bias}
\end{figure}


\begin{figure}
  \centering
  \includegraphics[width=.8\linewidth]{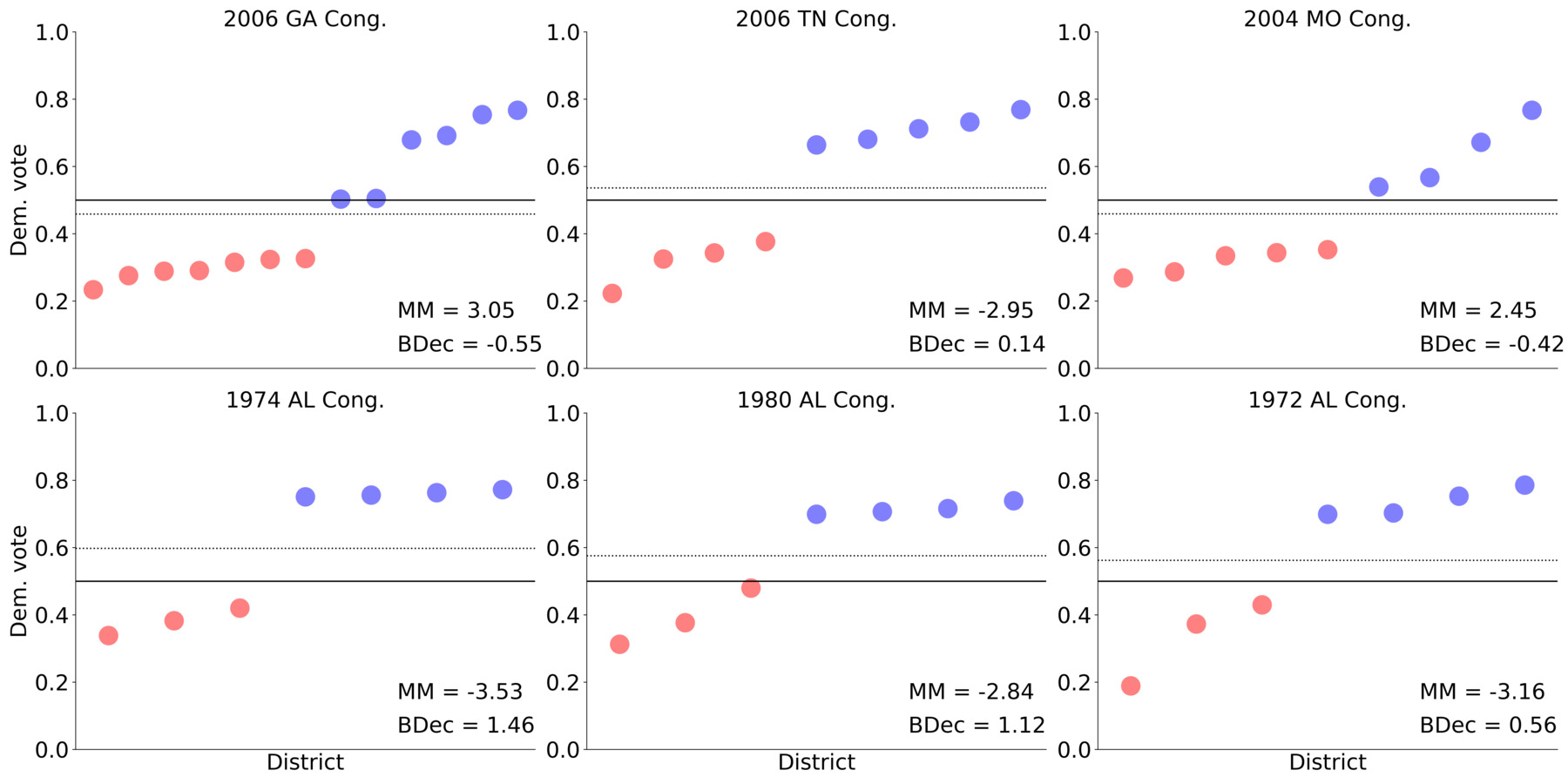}
  \caption{\mycapo}
  \label{fig:worst-BDec-MM}
\end{figure}


\begin{figure}
  \centering
  \includegraphics[width=.8\linewidth]{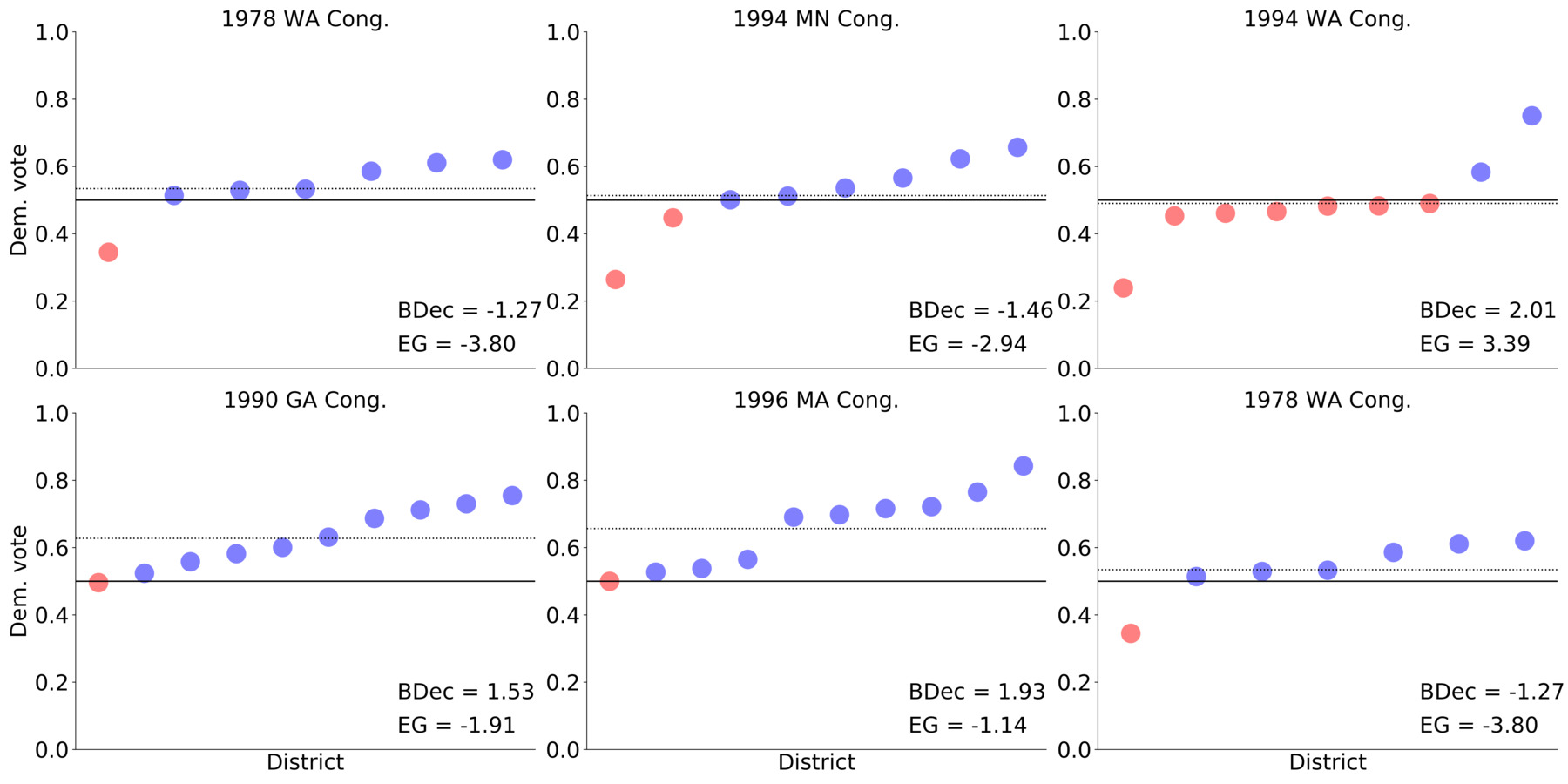}
  \caption{\mycapp}
  \label{fig:worst-EG-BDec}
\end{figure}


\begin{figure}
  \centering
  \includegraphics[width=.8\linewidth]{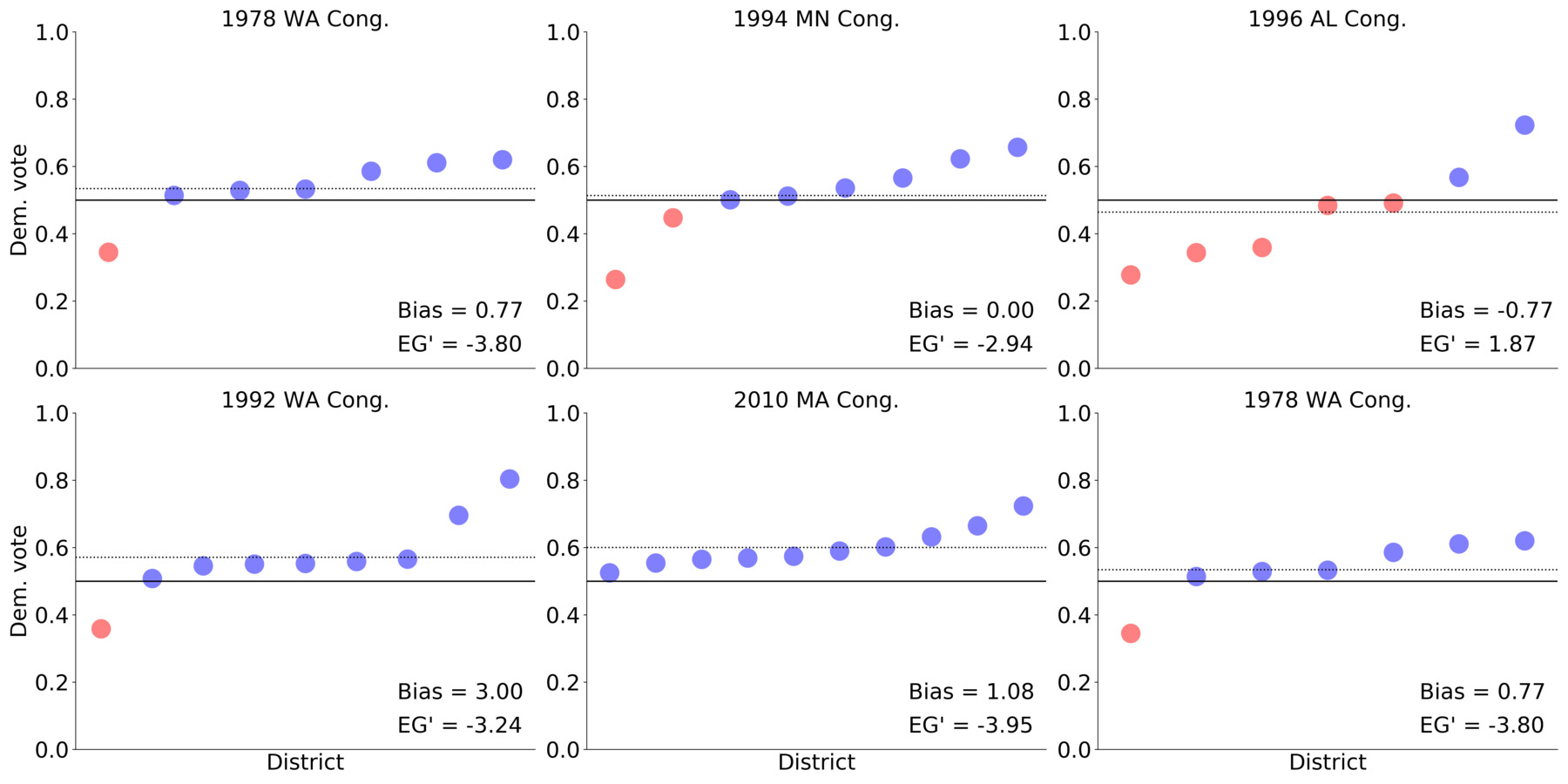}
  \caption{\mycapq}
  \label{fig:worst-EG-Bias}
\end{figure}


\begin{figure}
  \centering
  \includegraphics[width=.8\linewidth]{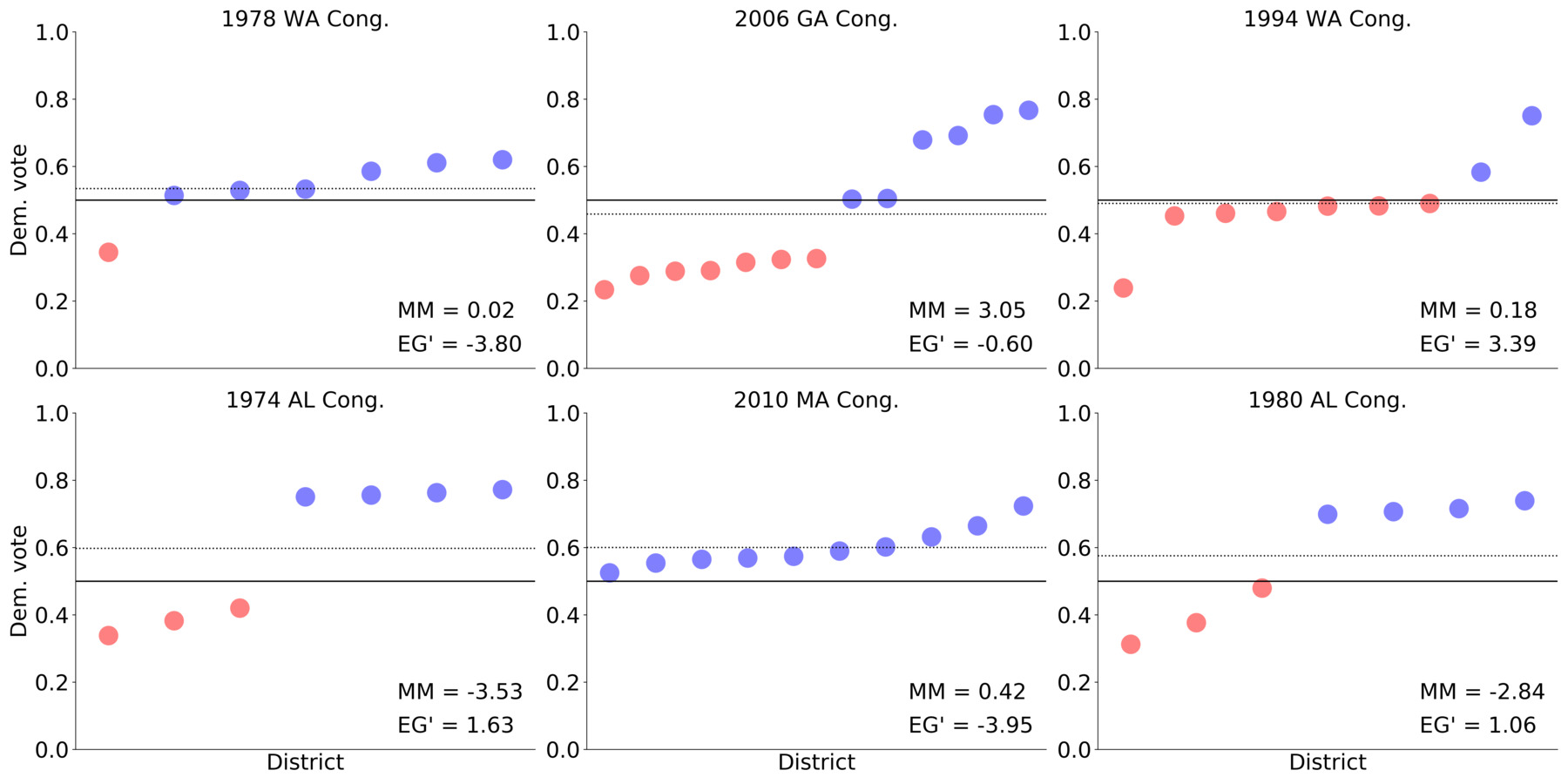}
  \caption{\mycapr}
  \label{fig:worst-EG-MM}
\end{figure}


\begin{figure}
  \centering
  \includegraphics[width=.8\linewidth]{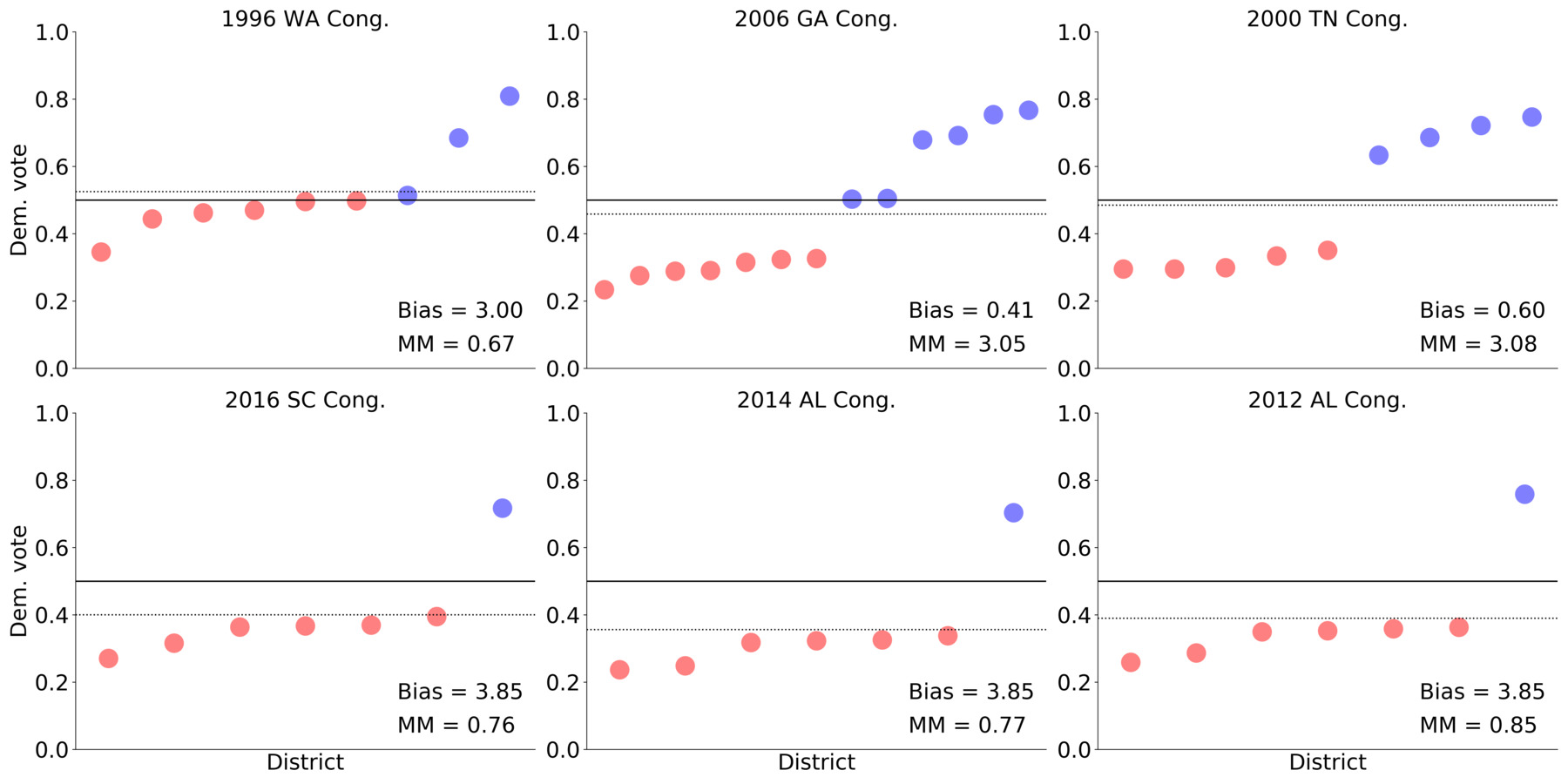}
  \caption{\mycaps}
  \label{fig:worst-MM-Bias}
\end{figure}


\clearpage

\ifdef{\SUBMIT}
{

  \setcounter{table}{0}
  \renewcommand{\thetable}{\arabic{table}}%
  \setcounter{figure}{0}
  \renewcommand{\thefigure}{\arabic{figure}}%

{
\small
\begin{table}
\centering
\caption{Summary of measures considered and their acronyms.}
\begin{tabular}{lp{5in}}
\toprule
Acronym & Measure\\
\midrule
$\noscale{\eg}$ & \emph{Efficiency gap:} comparison of wasted votes.\\
$\noscale{\dg}$ & \emph{Margin efficiency gap:} Variant of $\noscale{\eg}$ in which wasted votes are relative to losing party vote.\\
$\noscale{\lossg}$ & \emph{Losing efficiency gap:} Variant of $\noscale{\eg}$ in which only losing wasted votes are compared.\\
$\noscale{\va}$ & \emph{Vote-centric efficiency gap:} Variant of $\noscale{\eg}$ in which each comparison is between the fraction of votes wasted by each party rather than between the absolute amounts.\\
$\noscale{\vb}$ & \emph{Vote-centric losing efficiency gap:} Combination of $\noscale{\dg}$ and $\noscale{\va}$.\\
$\noscale{\taugap}$ & \emph{$\tau$-Gap:} Variant of $\noscale{\eg}$ in which votes are weighted according to a function parameterized by $\tau$.\\
$\noscale{\dec}$ & \emph{Declination:} A measure of differential responsiveness.\\
$\noscale{\bdec}$ & \emph{Buffered declination:} Variant of $\noscale{\dec}$ that is less sensitive in cases in which one party wins most of the votes.\\
$\noscale{\mm}$ & \emph{Mean-median difference:} A comparison of the median and mean of distribution.\\
$\noscale{\bias}$ & \emph{Partisan bias:} A comparison of seat share at 50\% of the statewide vote.\\
$\noscale{\biasO}$ & \emph{Specific asymmetry:} Variant of partisan bias that focuses on symmetry relative to observed vote.\\
$\noscale{\lm}$ &  \emph{Lopsided means:} A comparison of the average winning margins of parties.\\
$\noscale{\evw}$ & \emph{Equal vote weight:} Variant of $\noscale{\mm}$ that requires an anti-majoritarian outcome for a positive result.\\
\bottomrule
\label{tab:measures}
\end{tabular}
\end{table}
}

\clearpage

{
\small
\begin{table}
\centering
\caption{Mean and standard deviation for the measures considered in
  this article applied to the 1,166 elections in our data set with at
  least seven seats for which each party wins at least one seat.}
\begin{tabular}{lrrrrrrrrrrrrrr}
\toprule
{} &   $\noscale{\eg}$ &   $\noscale{\dg}$ &    $\noscale{\lossg}$ &   
   $\noscale{\va}$ &   $\noscale{\vb}$ & $\noscale{\taugap}$ 
&  $\noscale{\dec}$ &  $\noscale{\bdec}$ &    $\noscale{\mm}$ &
$\noscale{\bias}$ &  $\noscale{\biasO}$ & $\noscale{\lm}$ &   $\noscale{\evw}$\\
\midrule
Mean    & 0.01 &             0.04 &            -0.02 &              -0.05 &              -0.01 & -0.04 & -0.01 &  0.01 & -0.00 & -0.01 &               0.01 & 0.03 & -0.00 \\
Std Dev & 0.07 &             0.08 &             0.11 &               0.23 &               0.17 &  0.25 &  0.20 &  0.11 &  0.04 &  0.09 &               0.06 & 0.06 &  0.04 \\
\bottomrule
\label{tab:mean}
\end{tabular}
\end{table}
}

\clearpage

{\tiny
  \begin{table}
    \centering
    \caption{Outliers listed as last two digit of year along with
      state abbreviation. The first group of ten corresponds to the
      population of the 647 most balanced elections. The second
      group of ten corresponds to the population of all 1,179
      elections. In each grouping, the most extreme outliers are
      listed first. An asterisk denotes a state, rather than US House,
      election.}
\begin{tabular}{@{} cllllllllllllll @{}}
\toprule
& {} &       EG & $\mathrm{EG}^2$ & $\mathrm{EG}^0$ & $\mathrm{EG}_v^1$ & $\mathrm{EG}_v^2$ &  $\mathrm{EG}_\tau^1$ & Dec &    Dec' &      MM &    Bias & Bias' &     Lop &     EVW \\
\cmidrule{2-14}
& 1  &  78 WA  &          78 WA  &          78 WA  &            78 WA  &           78 WA  &  78 WA  &  78 WA  &  12 PA  &  00 TN  &  96 WA  &            14 NC  &  96 WA  &  80 IL  \\
& 2  &  94 WA  &          94 WA  &          94 WA  &            94 WA  &           94 WA  &  12 OH  &  12 OH  &  12 OH  &  06 GA  &  16 NC  &            04 AZ  &  80 WA  &  04 MI  \\
& 3  &  12 OH  &          06 MI  &          12 OH  &            12 OH  &           12 OH  &  12 PA  &  14 NC  &  16 PA  &  06 TN  &  12 NC  &            12 OH  &  94 WA  &  88 KY  \\
& 4  &  12 PA  &          96 WA  &          14 NC  &            12 IN  &           12 PA  &  14 NC  &  12 PA  &  14 NC  &  06 MO  &  14 NC  &            12 NC  &  94 AL  &  88 WI  \\
& 5  &  94 MN  &          12 PA  &          12 IN  &            14 NC  &           94 MN  &  94 WA  &  94 WA  &  12 MI  &  80 IL  &  12 OH  &            12 VA  &  98 AL  &  98 TX  \\
& 6  &  96 WA  &          12 MI  &          94 MN  &            94 MN  &           14 NC  &  16 NC  &  12 VA  &  06 MI  &  80 KY  &  04 AZ  &            12 PA  &  12 WI  &  84 WI  \\
& 7  &  12 NC  &          10 IL  &          16 NC  &            16 NC  &           12 VA  &  12 IN  &  16 NC  &  14 PA  &  06 AL  &  06 VA  &            96 WA  &  14 NJ  &  78 IL  \\
& 8  &  12 VA  &          12 OH  &          12 PA  &            12 PA  &           96 WA  &  16 PA  &  16 PA  &  92 TX  &  96 VA  &  12 VA  &            16 PA  &  94 VA  &  02 IL  \\ \rot{\rlap{Balanced elections}}
& 9  &  14 NC  &          12 NC  &          12 VA  &            12 VA  &           12 NC  &  94 MN  &  72 KY  &  12 NC  &  08 GA  &  12 PA  &            80 WA  &  84 TN  &  14 MI  \\
& 10  &  80 WA  &          06 OH  &         06 VA &            06 VA  &            12 IN  &  12 VA  &  06 VA  &  10 IL  &  04 MO  &  16 PA  &            72 KY  &  72 MN  &  16 MI  \\ \cmidrule{2-14}
& 1 &  10 MA  &          06 NY* &          10 MA  &            10 MA  &            10 MA  &  10 MA  &  80 VA  &  92 HI* &  74 AL  &  16 SC  &            14 NC  &  86 MD  &  80 IL  \\
& 2 &  78 WA  &          08 NY* &          14 MA  &            14 MA  &            78 WA  &  14 MA  &  90 MA  &  12 PA  &  72 AL  &  14 AL  &            04 AZ  &  82 AL  &  04 MI  \\
& 3 &  94 WA  &          06 NY  &          02 MA  &            02 MA  &            14 MA  &  02 MA  &  76 TX  &  12 OH  &  00 TN  &  12 AL  &            12 OH  &  74 MD  &  88 KY  \\
& 4 &  92 WA  &          00 NY* &          12 MA  &            12 MA  &            80 VA  &  80 VA  &  79 MS* &  78 TX* &  06 GA  &  86 MA  &            12 NC  &  92 MA  &  88 WI  \\
& 5 &  80 VA  &          16 NY  &          78 WA  &            16 MA  &            92 WA  &  12 MA  &  92 HI* &  76 TX  &  06 TN  &  92 WA  &            12 VA  &  98 MD  &  98 TX  \\
& 6 &  12 OH  &          96 NY* &          80 VA  &            04 MA  &            94 WA  &  78 WA  &  72 GA  &  94 HI* &  80 AL  &  14 IN  &            12 PA  &  74 NC  &  84 WI  \\
& 7 &  14 MA  &          98 NY* &          92 WA  &            76 GA  &            12 SC  &  79 MS* &  88 MA  &  16 PA  &  06 MO  &  96 WA  &            96 WA  &  96 MD  &  78 IL  \\ \rot{\rlap{All elections}}
& 8 &  12 PA  &          72 LA* &          79 MS* &            74 GA  &            02 MA  &  04 MA  &  12 SC  &  86 AL* &  86 WA  &  16 TN  &            16 PA  &  96 WA  &  02 IL  \\
& 9 &  12 SC  &          04 NY* &          16 MA  &            98 MA  &            12 MA  &  72 MO  &  84 MA  &  14 NC  &  86 FL  &  16 NC  &            72 KY  &  74 MN  &  14 MI  \\
& 10 &  94 MN  &          78 WA  &         04 MA  &            08 MA  &            12 OH  &  12 SC  &  86 MA  &  76 TX* &  80 IL  &  14 NC  &            98 AL  &  72 WI  &  16 MI  \\
\bottomrule
\end{tabular}
\label{tab:outliers}
\end{table}
}

\clearpage

\begin{table}
  \centering
\caption{Values of scaled measures on hypothetical elections from
  Figure~\ref{fig:hypo_grid}. Column $N$ provides the number of
  districts in each election while the statewide Democratic vote share
  is indicated in the Mean column. Bold entries indicate false
  positives/negatives.}
\begin{tabular}{lllllllllllllll}
\toprule
{} &   N &  Mean &             $\scale{\eg}$ &             $\scale{\dg}$ &             $\scale{\lossg}$ &     $\scale{\va}$ &      $\scale{\vb}$ &  $\scale{\taugap}$  &     $\scale{\dec}$ &           $\scale{\bdec}$ &         $\scale{\mm}$ &           $\scale{\bias}$ &          $\scale{\biasO}$ &            $\scale{\lm}$ \\
\midrule
\elecA  &  23 &  0.60 &            \phantom{-}1.2 &    \phantom{-}\textbf{2.3} &            -0.1 &              -0.1 &               \phantom{-}0.6 &                    \phantom{-}0.5 &            \phantom{-}0.1 &            \phantom{-}0.7 &            \phantom{-}0.0 &            \phantom{-}0.2 &              \phantom{-}0.0 &   \phantom{-}\textbf{2.0} \\
\elecB  &  25 &  0.60 &            \phantom{-}0.3 &             \phantom{-}1.5 &            -0.7 &              -0.7 &              -0.1 &                   -0.6 &            \phantom{-}0.1 &            \phantom{-}0.8 &            \phantom{-}0.0 &            \phantom{-}0.2 &              \phantom{-}0.0 &            \phantom{-}1.9 \\
\elecC  &  25 &  0.60 &           -1.3 &             \phantom{-}0.0 &            -1.8 &              -1.8 &              -1.5 &                   -1.8 &            \phantom{-}0.0 &            \phantom{-}0.8 &            \phantom{-}0.0 &            \phantom{-}0.2 &              \phantom{-}0.0 &            \phantom{-}1.8 \\
\elecD  &  10 &  0.64 &  \textbf{-2.9} &            -1.0 &   \textbf{-3.3} &     \textbf{-3.4} &     \textbf{-3.3} &          \textbf{-3.3} &             &             &            \phantom{-}0.0 &            \phantom{-}0.0 &              \phantom{-}0.0 &             \\\midrule
\elecE  &  12 &  0.52 &  \textbf{-3.9} &   \textbf{-3.4} &   \textbf{-2.8} &     \textbf{-2.7} &     \textbf{-3.5} &          \textbf{-2.7} &           -0.1 &            \phantom{-}0.1 &            \phantom{-}0.3 &            \phantom{-}0.9 &    \textbf{-2.0} &            \phantom{-}0.3 \\
\elecF  &  10 &  0.51 &  \textbf{-2.4} &   \textbf{-2.1} &            -1.7 &              -1.6 &     \textbf{-2.1} &                   -1.6 &           -1.0 &           -0.7 &           -0.1 &  \textbf{-2.2} &              \phantom{-}0.0 &           -0.5 \\
\elecG  &  10 &  0.52 &           -0.7 &            -0.4 &            -0.7 &              -0.7 &              -0.7 &                   -0.7 &           -0.8 &           -0.8 &  \textbf{-2.1} &           -1.1 &             -1.6 &           -0.5 \\
\elecH  &  10 &  0.52 &           -0.7 &            -0.4 &            -0.7 &              -0.7 &              -0.8 &                   -0.9 &           -1.0 &           -1.2 &  \textbf{-4.4} &           -1.1 &             -1.6 &           -1.4 \\\midrule

\elecI  &  10 &  0.57 &  -2.1 &            \textbf{-1.1} &   -2.1 &     -2.0 &     -2.2 &                   \textbf{-1.8} &  -2.5 &  -2.0 &            \phantom{-}\textbf{0.0} &            \phantom{-}\textbf{0.0} &              \phantom{-}\textbf{0.0} &           \textbf{-1.5} \\
\elecJ  &  10 &  0.44 &  -2.9 &   -3.3 &            \textbf{-1.4} &              \textbf{-1.4} &     -2.3 &                   \textbf{-1.6} &  -2.0 &  -2.7 &  -2.1 &           \textbf{-1.1} &             \textbf{-1.6} &  -3.4 \\
\elecK &  10 &  0.50 &   \phantom{-}2.7 &    \phantom{-}2.5 &             \phantom{-}\textbf{1.8} &               \phantom{-}\textbf{1.7} &      \phantom{-}2.4 &                    \phantom{-}\textbf{1.8} &   \phantom{-}2.2 &   \phantom{-}2.2 &            \phantom{-}\textbf{1.4} &   \phantom{-}2.2 &     \phantom{-}3.2 &   \phantom{-}2.2 \\
\elecL &  10 &  0.30 & -2.7 &   -4.9 &             \phantom{-}\textbf{0.0} &               \phantom{-}\textbf{0.0} &              \textbf{-1.6} &                   \textbf{-0.3} &  -2.0 &  -3.2 &            \phantom{-}\textbf{1.4} &   \phantom{-}\textbf{2.2} &    -2.4 &  -5.0 \\
\bottomrule
\end{tabular}
\label{tab:hypo}
\end{table}



\clearpage

\begin{figure}
  \centering
  \includegraphics[width=1\linewidth]{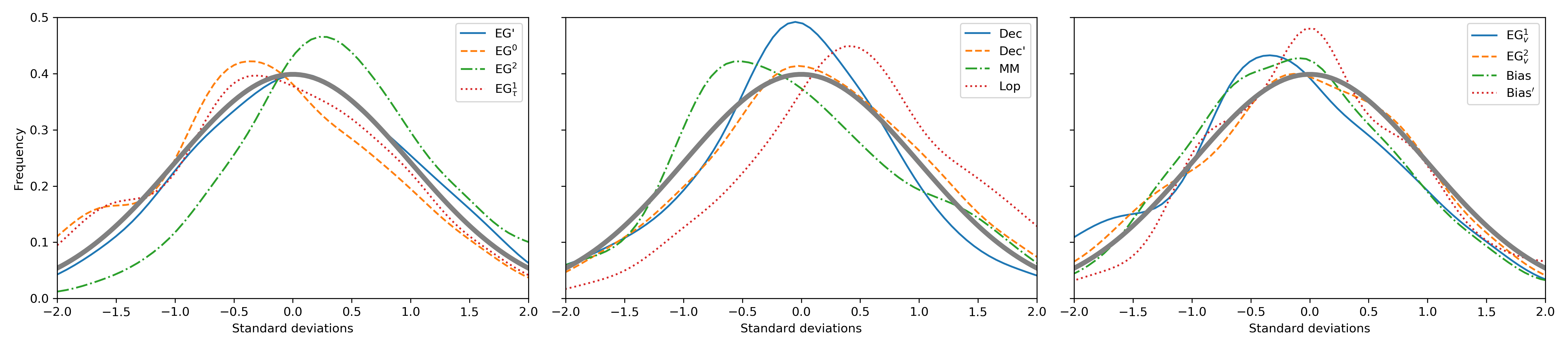}
  \caption{\mycapc}
  \label{fig:risk-kde}
\end{figure}

\clearpage

\begin{figure}
  \centering
  \includegraphics[width=.6\linewidth]{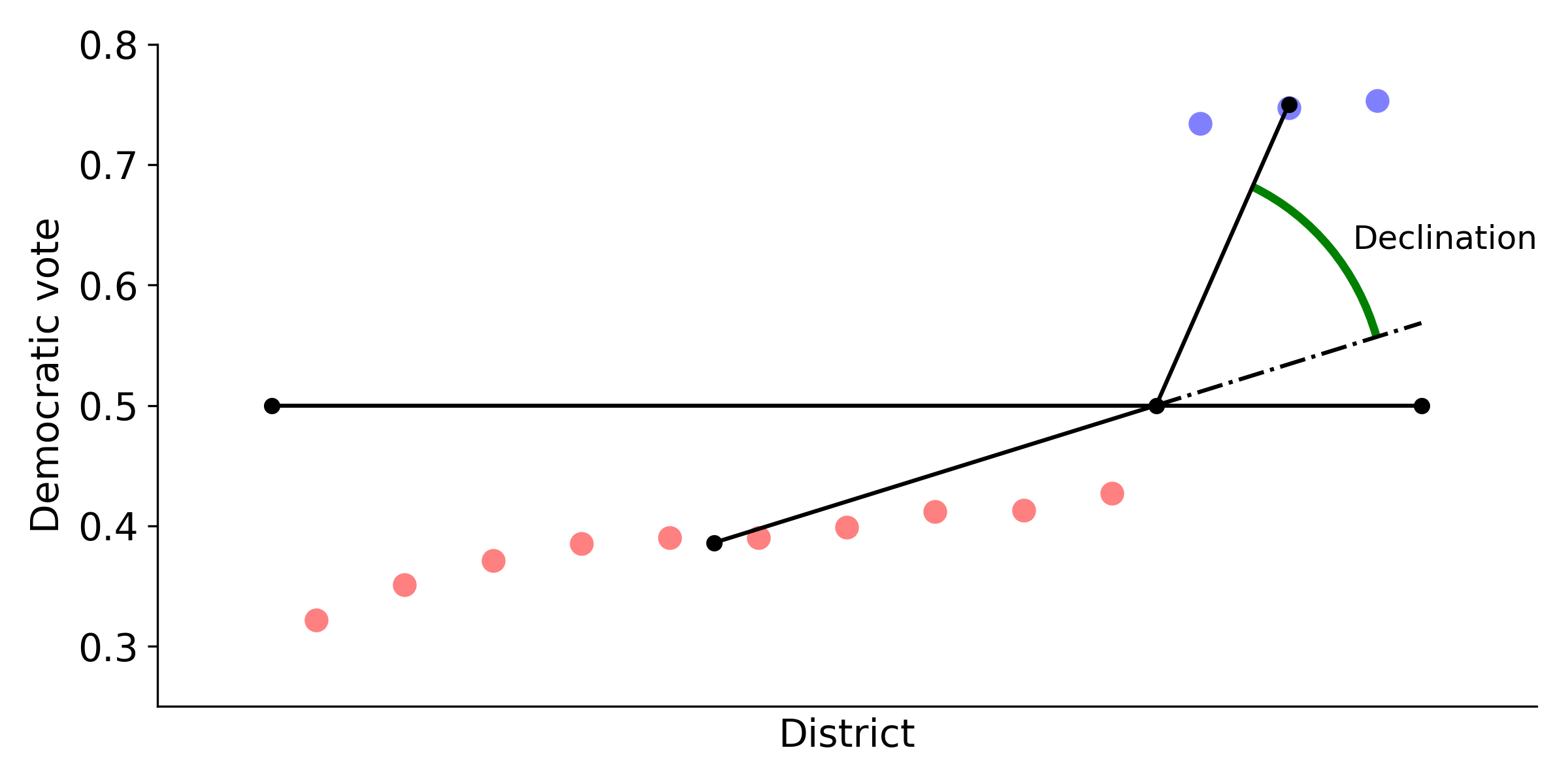}
  \caption{\mycapb}
  \label{fig:plot}
\end{figure}

\clearpage

\begin{figure}
  \centering
  \includegraphics[width=1\linewidth]{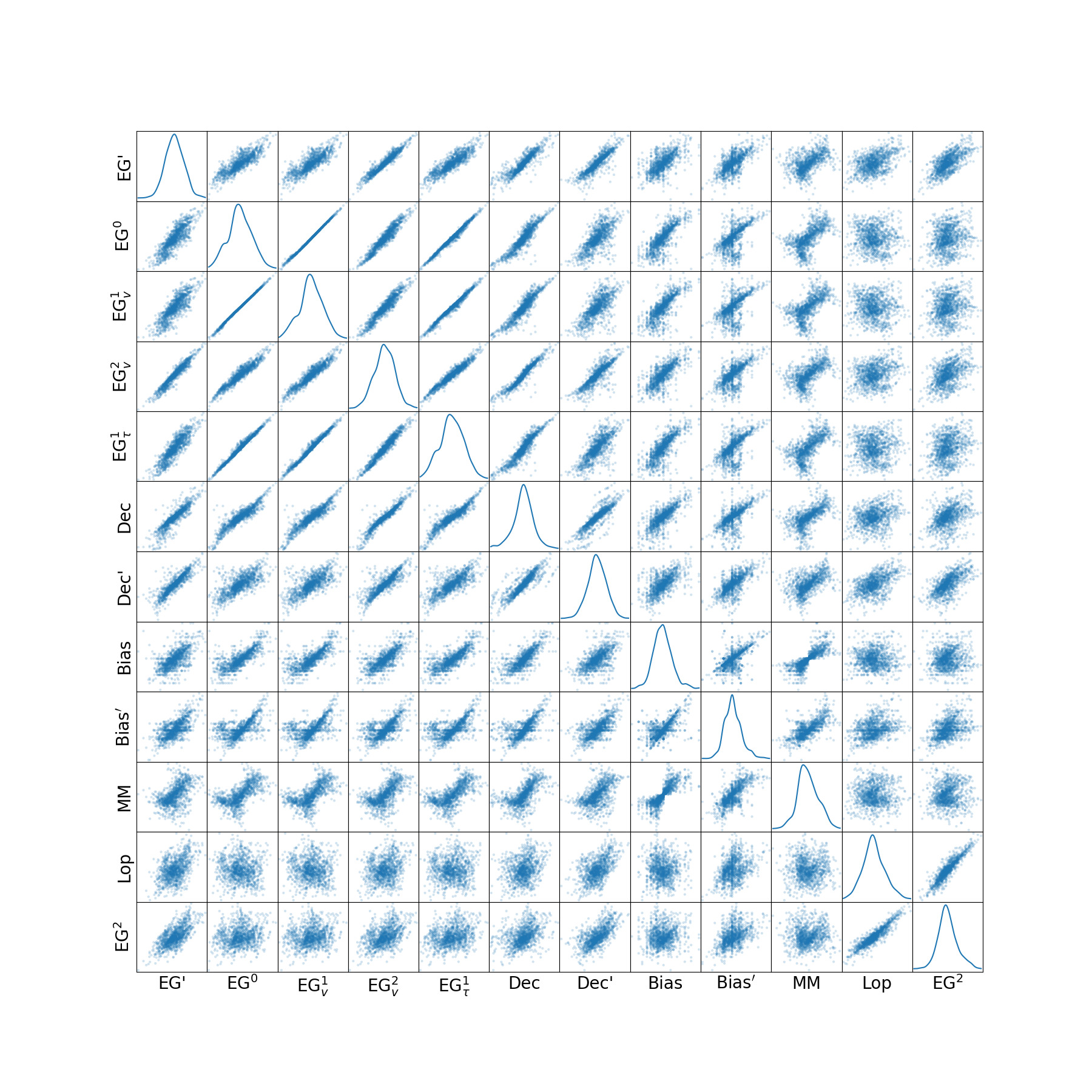}
  \caption{\mycapd}
  \label{fig:measure-scatter}
\end{figure}

\clearpage

\begin{figure}
  \centering
  \includegraphics[width=.9\linewidth]{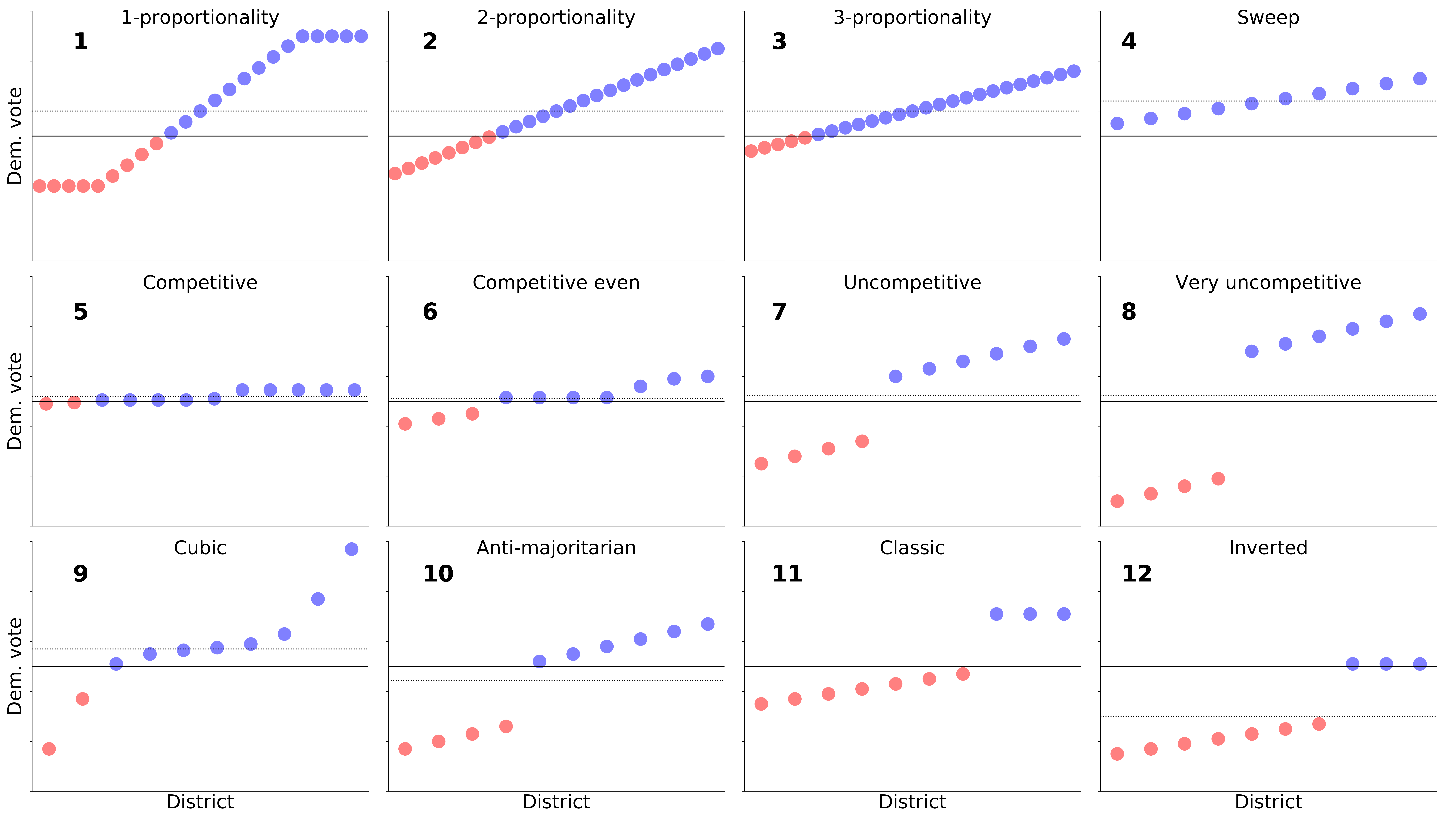}
  \caption{\mycapm}
  \label{fig:hypo_grid}
\end{figure}

\clearpage

\begin{figure}
  \centering
  \includegraphics[width=1\linewidth]{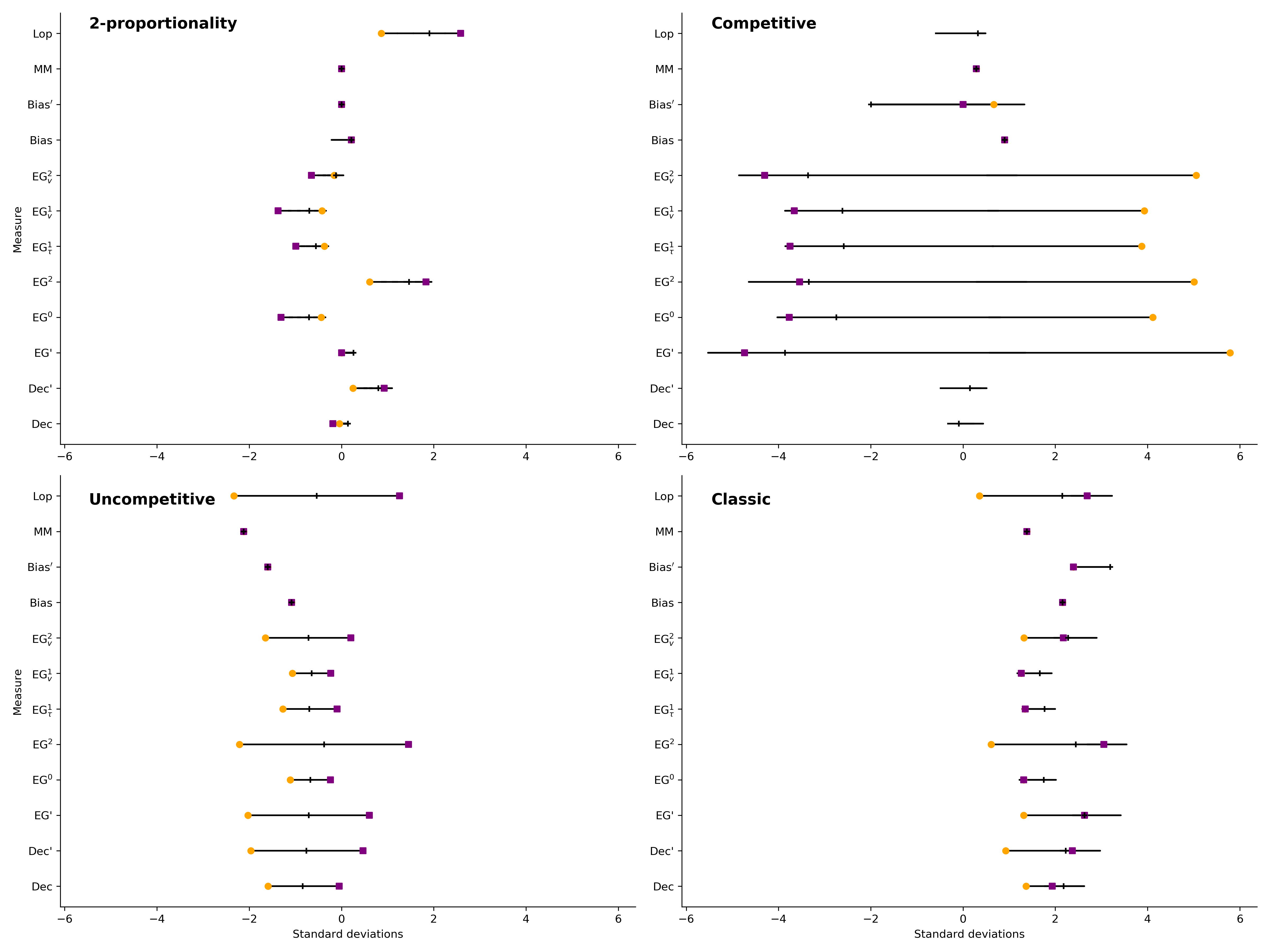}
  \caption{\mycapsens}
  \label{fig:sens}
\end{figure}

\clearpage

\pagestyle{empty}

\clearpage
{\bf Figure 1: }\mycapb
\clearpage
{\bf Figure 2: }\mycapc
\clearpage
{\bf Figure 3: }\mycapd
\clearpage
{\bf Figure 4: }\mycapm
\clearpage
{\bf Figure 5: }\mycapsens

}{}

\end{document}